\documentclass[rmp,aps,nofootinbib,eqsecnum,twocolumn]{revtex4}

\usepackage{graphicx}
\usepackage{amsmath,amssymb,amsthm}
\usepackage{bbm}
\usepackage{psfrag}
\usepackage[dvipsnames]{xcolor}
\usepackage{ctable}

\newcommand\NNN{{\mathbbm{N}}}
\newcommand\ZZ{{\mathbbm{Z}}}
\newcommand\RR{{\mathbbm{R}}}
\newcommand\ee{{\mathrm{e}}}
\newcommand\dd{{\mathrm{d}}}
\newcommand\MM{{\mathcal{M}}}

\DeclareMathOperator{\sgn}{sgn}
\DeclareMathOperator{\Order}{{\mathcal O}}
\DeclareMathOperator{\smallo}{o}
\DeclareMathOperator{\Heaviside}{\Theta}
\DeclareMathOperator{\Dirac}{\delta}

\theoremstyle{definition}
\newtheorem{defn}{Definition}[section]
\newtheorem{thrm}[defn]{Theorem}
\newtheorem{conjecture}[defn]{Conjecture}

\begin{document}
\title{Phase transitions and configuration space topology}

\author{Michael Kastner}
\email{Michael.Kastner@uni-bayreuth.de}
\affiliation{Physikalisches Institut, Universit\"at Bayreuth, 95440 Bayreuth, Germany}

\begin{abstract}
Equilibrium phase transitions may be defined as nonanalytic points of thermodynamic functions, e.\,g., of the canonical free energy. Given a certain physical system, it is of interest to understand which properties of the system account for the presence, or the absence, of a phase transition, and an investigation of these properties may lead to a deeper understanding of the physical phenomenon. One possible way to approach this problem, reviewed and discussed in the present paper, is the study of topology changes in configuration space which, remarkably, are found to be related to equilibrium phase transitions in classical statistical mechanical systems. For the study of configuration space topology, one considers the subsets $\MM_v$, consisting of all points from configuration space with a potential energy per particle equal to or less than a given $v$. For finite systems, topology changes of $\MM_v$ are intimately related to nonanalytic points of the microcanonical entropy (which, as a surprise to many, do exist). In the thermodynamic limit, a more complex relation between nonanalytic points of thermodynamic functions (i.\,e., phase transitions) and topology changes is observed. For some class of short-range systems, a topology change of the $\MM_v$ at $v=v_{\text{t}}$ was proven to be necessary, but not sufficient, for a phase transition to take place at a potential energy $v_{\text{t}}$. In contrast, phase transitions in systems with long-range interactions or in systems with nonconfining potentials need not be accompanied by such a topology change. Instead, for such systems the nonanalytic point in a thermodynamic function is found to have some maximization procedure at its origin. These results may foster insight into the mechanisms which lead to the occurrence of a phase transition, and thus may help to explore the origin of this physical phenomenon.
\end{abstract}

\date{October 08, 2007}


\maketitle

\tableofcontents

\section*{Preface}
It was at the end of the 1990s when the application of concepts from differential geometry to Hamiltonian dynamical systems led to a conjectured connection between the occurrence of equilibrium phase transitions in classical Hamiltonian systems and some topological quantities of configuration space subsets of these systems. Since then, the interest in this approach and the number of people working on the topic has increased, and so has the number of results.

At the time of this writing, an overview of the subject is difficult to attain. First, the results are scattered among a considerable number of publications and, second, several of the results, although correct, demand a reinterpretation as a consequence of recent findings and developments. The purpose of the present paper is to assemble from the known results, as far as possible, a coherent picture of the relation between phase transitions and configuration space topology, and to indicate new lines of research which might open up from these concepts.

\section{Introduction}
Phase transitions, like the boiling and evaporating of water at a certain temperature and pressure, are common phenomena both in everyday life and in almost any branch of physics. Loosely speaking, a phase transition brings about a sudden change of the macroscopic properties of a many-particle system while smoothly varying a parameter (the temperature or the pressure in the above example). Probably the main reason for the unabated interest that phase transitions have received already for more than a century is their omnipresence in all branches of physics (and also in related fields like biology or engineering): be it the formation of stars in astrophysics, the transition to superconductivity in solid state physics, or the opening of the DNA helix in biophysics, examples of many-particle systems which undergo a phase transition are widespread and of indisputable relevance in science.

Phase transitions can occur in both equilibrium and nonequilibrium systems, but the focus will be exclusively on the equilibrium case in this exposition. There is a plethora of books on the subject, ranging from experimental to theoretical and mathematical treatises. Especially the theory books are to a large extent concerned with the classification of different types of phase transition: remarkably, very different physical systems may show quantitatively identical properties in the vicinity of the phase transition point, and this fascinating topic of {\em universality}, explained using the renormalization group theory, has attracted a lot of interest especially in the 1970s and 1980s \cite{Binney_etal,Lesne}. In the present paper, instead of discussing the characteristics of the different types of phase transition, we take one step back and inquire about the conditions which may lead to the occurrence of a phase transition in a given system.

The mathematical description of equilibrium phase transitions is conventionally based either on Gibbs measures on phase space or on (grand)canonical thermodynamic functions, relating their loss of analyticity (or, in other words, the appearance of a singularity) to the occurrence of a phase transition.\footnote{Standard references in mathematical physics for these two points of view are by \textcite{Ruelle} and \textcite{Georgii}.} A nonanalyticity of a thermodynamic function leads to a discontinuity or to a divergence in some derivative of this function, and this is a typical signature of a phase transition as measured experimentally. Within the (grand)canonical ensemble of statistical mechanics, such a nonanalytic behavior can occur only in the thermodynamic limit, in which the number of degrees of freedom $N$ of the system goes to infinity.\footnote{Apparently it was first suggested by H.\ A.\ Kramers in 1937 to take this limit.} Conceptually, the necessity of the thermodynamic limit is an objectionable feature: first, the number of degrees of freedom in real systems, although possibly large, is finite, and, second, for systems with long-range interactions, the thermodynamic limit may even be not well defined. These observations indicate that the theoretical description of phase transitions, although very successful in certain aspects, may not be completely satisfactory.

Apart from this conceptual shortcoming, in the field of phase transitions there are many problems of applied nature which are far from being settled. One of those is the search for sufficient or necessary conditions for the occurrence of a phase transition. Among the {\em necessary}\/ conditions for the occurrence of a phase transition, there are some of reasonable generality, like the Mermin-Wagner theorem and its generalizations \cite{MerWag:66,FroePfi:81,FanVanVer:84} or the theorems on the absence of phase transitions in certain one-dimensional systems by \textcite{vanHove:50} and by \textcite{CuSa:04}. Yet improved criteria are of course desirable. Less is known about conditions {\em sufficient}\/ to guarantee a phase transition to take place. The Peierls argument \cite{Peierls:36} or the Fr\"ohlich-Simon-Spencer bound \cite{FroeSiSpe:76} can be used to prove the existence of phase transitions without explicitly computing a thermodynamic potential, but their application is model specific and may be difficult depending on the system of interest.

The above considerations motivate a further study on the ``nature'' of phase transitions, of the underlying mechanisms leading to a nonanalytic point of a thermodynamic function, and of the conditions under which they can occur. A classic result identifying such a nonanalyticity generating mechanism is the seminal theorem of Lee and Yang, relating the properties of the zeros of the grandcanonical partition function in the complex fugacity plane to nonanalyticities of the corresponding thermodynamic function \cite{LeeYang:52}. The main issue of the present paper is to investigate the mechanism which is at the basis of a phase transition using a different approach, based on concepts from differential geometry and topology.

This {\em topological approach}\/ emerged from the study of Hamiltonian dynamical systems and is therefore---at least in its present formulation---applicable to classical (i.\,e., nonquantum mechanical) systems. Hamiltonian dynamics can be viewed as a geo\-de\-sic flow on the configuration space, provided the latter is equipped with a suitable metric.\footnote{Both Hamiltonian dynamics and geodesics of a Riemannian manifold can be defined by some variational principle: the trajectories of Hamiltonian dynamics are the extrema of the Hamiltonian action functional, whereas the geodesics of any Riemannian manifold are given by the extrema of the arclength functional. It is this structural similarity which allows for a geometric formulation of Hamiltonian dynamics [see \textcite{MarsRa} or \textcite{CaPeCo:00}].} By numerical methods, geometric quantities of such metric spaces of Hamiltonian dynamical systems were studied as a function of energy or temperature in a series of papers [see \textcite{CaPeCo:00} and \textcite{Pettini} for reviews]. For a system undergoing a phase transition in the thermodynamic limit, these geometric quantities display discontinuous or cusp-like features remarkably close to the transition energy or temperature. One possible mechanism behind such a dramatic change of {\em geometric}\/ quantities of configuration space can be a change of its {\em topology}, and it was this line of reasoning which led Pettini and co-workers to conjecture a connection between phase transitions, on the one hand, and topology changes within a family of certain configuration space subsets, on the other hand \cite{CaCaClePe:97}. Subsequently, such a connection was proven to hold true for a certain class of systems with short-range interactions, showing that a topology change is a {\em necessary}\/ condition for a phase transition to take place \cite{FraPe:04,FraPeSpi,FraPe}. This theorem, for the class of systems covered by its assumptions, suggests the interpretation of certain topology changes as the relevant mechanism behind the generation of a nonanalyticity in a thermodynamic function, and it furthermore allows one to exclude the occurrence of a phase transition when such topology changes are absent.

The use of concepts from topology to describe a physical phenomenon is particularly appealing due to the fact that topology yields a very reductional description: considering only the topology of, say, a surface, a significant amount of ``information'' (on curvatures, for example) is disregarded, and only a small part (like connectivity properties) is kept. If one then succeeds to capture the essentials of the phenomenon of interest with the remaining information only, the resulting description will be an efficient one, and one might hope to get an unblurred view onto the mechanism which is at the basis of the phenomenon.

The initial hope that such a topological approach might provide a unified and completely general description of phase transitions turned out to be over-optimistic. It was only in the last few years that evidence accumulated, disproving the general validity of the hypothesized connection between phase transitions and configuration space topology \cite{BaroniPhD,GaSchiSca:04,AnRuZa:05,HaKa:05}. This observation, perceived as a major set-back at first, appears less dramatic in light of subsequent findings. However, and this was one of the motivations for writing the present paper, it alters the understanding of the topological approach to phase transitions, as well as of several of the results of model calculations reported in the literature. The aim of the present paper is to review results on the relation between configuration space topology and analyticity properties of thermodynamic functions. This relation is found to depend on the physical situation, in particular on whether the system of interest is finite or infinite and whether the interparticle interactions are of short range or of long range. The results may help to deepen the understanding of the basic mechanisms behind phase transitions in the infinite system case, and they can explain the peculiar (non)analyticity properties of microcanonical entropy functions of finite systems. 

The paper is structured as follows: We start by fixing notations and giving some basic definitions used in the topological approach to phase transitions in Sec.~\ref{sec:definitions}. Since Morse theory provides a suitable mathematical framework for the study of the topology of the configuration space subsets of interest, a summary of elementary results of this theory as well as an application to a statistical mechanical model is given in Sec.~\ref{sec:models}. The relation between certain topology changes and nonanalyticities in thermodynamic functions of {\em finite}\/ systems is discussed in Sec.~\ref{sec:finite_systems}. The rest of the paper is devoted to the more intricate case of {\em infinite}\/ systems. A theorem rigorously establishing a connection between phase transitions and topology changes in configuration space is presented in Sec.~\ref{sec:theorems}, where also results on the configuration space topology of models beyond this theorem's assumptions are reviewed. In Sec.~\ref{sec:limitations}, the limitations of the topological approach are explored by studying two models for which the proposed relation between phase transitions and configuration space topology turns out to be invalid. Some proposed sufficiency conditions on the topology changes, guaranteeing the occurrence of a phase transition, are critically discussed in Sec.~\ref{sec:sufficiency}. We conclude the paper with a summary in Sec.~\ref{sec:conclusions}.

\section{Definitions and preliminaries}
\label{sec:definitions}
Discussing topology and topology changes in the Introduction, vague reference was made to certain subsets of configuration space. It is the aim of the present section to fix notations and to define configuration space subsets and thermodynamic quantities to which reference will be made throughout.

\subsection{Standard Hamiltonian systems}
\label{sec:standard_H}
We consider classical Hamiltonian systems consisting of $N$ degrees of freedom, characterized by some Hamiltonian function
\begin{equation}
H:\Lambda_N\to\RR
\end{equation}
which maps the phase space $\Lambda_N\subseteq\RR^{2N}$ onto the reals. For convenience we assume $H$ to be of the standard form
\begin{equation}\label{eq:H_standard}
H(p;q)=\tfrac{1}{2}(p,p)+V(q),
\end{equation}
where $p=\left(p_1,\dotsc,p_N\right)$ is the vector of momenta and $q=\left(q_1,\dotsc,q_N\right)$ is the vector of position coordinates. Masses have been set to unity and $(\cdot,\cdot)$ denotes the usual scalar product. The potential
\begin{equation}\label{eq:V}
V:\Gamma_N\to\RR
\end{equation}
maps the {\em configuration space}\/ $\Gamma_N\subseteq\RR^{N}$ onto the reals. The restriction to Hamiltonian functions of the standard form \eqref{eq:H_standard} is convenient, but not essential for the concepts discussed.\footnote{For the thermodynamics of phase transitions, a quadratic form in the momenta as in Eq.\ \eqref{eq:H_standard} only leads to a shift in the free energy [however, see \textcite{CaKa:07} for a discussion of the pitfalls of this reasoning]. Furthermore, after becoming familiar with the topological concepts and notations, the reader may convince himself that any critical point $q_\text{c}$ of the potential $V$ corresponds to a critical point $(0;q_\text{c})$ of a standard Hamiltonian function, and that therefore the contribution of a standard kinetic energy as in Eq.\ \eqref{eq:H_standard} to the topological approach is a trivial one. As a consequence, the results presented remain valid also for models without a kinetic energy term (which is the situation typically encountered when studying spin systems like the Heisenberg model).}
Throughout the paper, we assume $\Gamma_N$ to be continuous (in contrast to classical spin models like the Ising model or the Potts model which have a discrete configuration space). This will be important for the kind of configuration space topology we consider in the following.

\subsection{Configuration space subsets}
\label{sec:subsets}
From the potential function $V$, we define the family of subsets $\left\{\MM_v\right\}_{v\in\RR}$, where
\begin{equation}\label{eq:Mv}
\MM_v=V^{-1}\left(-\infty,vN\right]=\left\{q\in\Gamma_N\,\big|\,V(q)\leqslant vN\right\}.
\end{equation}
$V^{-1}$ gives the preimage of a set under $V$, hence $\MM_v$ is the subset of all points $q$ from configuration space $\Gamma_N$ for which the potential energy per degree of freedom $V(q)/N$ is equal to or less than a given value $v$. Similarly, the related family $\left\{\Sigma_v\right\}_{v\in\RR}$ can be defined, where
\begin{equation}
\Sigma_v=V^{-1}\left(vN\right)=\left\{q\in\Gamma_N\,\big|\,V(q)= vN\right\}
\end{equation}
consists of all points $q$ from the configuration space $\Gamma_N$ for which the potential energy per degree of freedom $V(q)/N$ is equal to a given value $v$. These constant potential energy subsets form the boundaries of the corresponding $\MM_v$, i.\,e.,
\begin{equation}
\Sigma_v=\partial\MM_v,
\end{equation}
so that $\MM_v$ or $\Sigma_v$ are closely related. It is therefore a matter of convenience to use one quantity or the other, depending on the actual situation of interest. The main topic of this paper is the relation of topology changes of $\MM_v$ or $\Sigma_v$ to nonanalytic points of thermodynamic functions.

\subsection{Thermodynamic functions}
\label{sec:td_functions}
Strictly speaking, the notion of a thermodynamic function should be restricted to functions describing the equilibrium behavior of systems in the thermodynamic limit of infinitely many degrees of freedom. For matters of convenience, we likewise will speak of thermodynamic functions when referring to their finite system counterparts.
 
Regarding their analyticity properties, thermodynamic functions obtained from different statistical ensembles can differ drastically. In this paper we discuss analyticity properties of microcanonical and canonical thermodynamic functions.\footnote{The extension to further statistical ensembles is straightforward.}

\subsubsection{Microcanonical thermodynamic functions}
The microcanonical ensemble provides the framework for the statistical description of an isolated physical system in which the total energy is conserved. The fundamental quantity of this ensemble is the {\em Boltzmann entropy}\/ or {\em microcanonical entropy}\/ as a function of the energy (per degree of freedom) $\varepsilon$,
\begin{equation}\label{eq:s}
\bar{s}_N(\varepsilon)=\frac{1}{N}\ln \int_{\Lambda_N} \dd p\, \dd q \Dirac\left[H(p;q)-N\varepsilon\right],
\end{equation}
where $\delta$ denotes the Dirac distribution.\footnote{Here and in the following we define thermodynamic functions always per degree of freedom, which accounts for the factor $1/N$ in the definitions.} A related quantity is the {\em configurational microcanonical entropy}\/ as a function of the potential energy (per degree of freedom) $v$,
\begin{equation}\label{eq:s_conf}
s_N(v)=\frac{1}{N}\ln \int_{\Gamma_N} \dd q \Dirac\left[V(q)-Nv\right].
\end{equation}

\subsubsection{Canonical thermodynamic functions}
The canonical ensemble provides the framework for the statistical description of a system coupled to an infinitely large heat bath of inverse temperature $\beta$. The fundamental quantity of this ensemble is the {\em canonical free energy}\/
\begin{equation}\label{eq:f}
\bar{f}_N(\beta)=-\frac{1}{N\beta}\ln \int_{\Lambda_N} \dd p\, \dd q\, \ee^{-\beta H(p;q)}. 
\end{equation}
A related quantity is the {\em configurational canonical free energy}\/
\begin{equation}\label{eq:f_conf}
f_N(\beta)=-\frac{1}{N\beta}\ln \int_{\Gamma_N} \dd q\, \ee^{-\beta V(q)}. 
\end{equation}

\subsubsection{Relation of microcanonical and canonical thermodynamic functions}
It follows from the definitions \eqref{eq:s} and \eqref{eq:f} that microcanonical entropy and canonical free energy are related by
\begin{equation}\label{eq:fbar_to_sbar}
\bar{f}_N(\beta)=-\frac{1}{N\beta}\ln \int_\RR \dd(N\varepsilon)\, \exp\left\{N\left[\bar{s}_N(\varepsilon)-\beta \varepsilon\right]\right\}.
\end{equation}
Similarly, for the configurational thermodynamic functions the relation
\begin{equation}\label{eq:f_to_s}
f_N(\beta)=-\frac{1}{N\beta}\ln \int_\RR \dd (Nv)\, \exp\left\{N\left[s_N(v)-\beta v\right]\right\}
\end{equation}
holds. In the thermodynamic limit, using Laplace's method for the evaluation of asymptotic integrals \cite{BenOrs}, well-known relations of thermodynamic functions by means of Legendre-Fenchel transformations,
\begin{equation}
-\beta\bar{f}_\infty(\beta) = \sup_\varepsilon \left[\bar{s}_\infty(\varepsilon)-\beta\varepsilon\right]
\end{equation}
and
\begin{equation}\label{eq:Legendre}
-\beta f_\infty(\beta) = \sup_v \left[s_\infty(v)-\beta v\right],
\end{equation}
are obtained from Eqs.\ \eqref{eq:fbar_to_sbar} and \eqref{eq:f_to_s}.

\subsection{Nonanalytic points and phase transitions}
The main topic of the present paper is the relation between topology changes within the families $\left\{\MM_v\right\}_{v\in\RR}$ or $\left\{\Sigma_v\right\}_{v\in\RR}$ of configuration space subsets defined in Sec.~\ref{sec:subsets}, on the one hand, and nonanalytic points of thermodynamic functions, on the other hand.

\begin{defn}
A {\em nonanalytic points}\/ is a point in the interior of the domain of a real function at which the function is not infinitely many times real-differentiable. Synonymously we speak of {\em nonanalyticities}\/ of the function.
\end{defn}

Different definitions of thermodynamic phase transitions can be found in the literature, where the most common ones are based either on the (non)\-unique\-ness of translationally invariant Gibbs measures on phase space or on the (non)analyticity of thermodynamic functions.\footnote{For some examples like the Ising model both definitions are known to coincide, for many others coincidence may be expected. However, counterexamples can be found as well, like the case of the Kosterlitz-Thouless phase transition of the two-dimensional $XY$ model which has a unique translationally invariant Gibbs state for all temperatures \cite{BricFonLan:77}. An introductory discussion has been given by \textcite{Lebowitz:99}.} Throughout this paper we will use the following version of the latter approach.
\begin{defn}\label{def:PT}
A {\em phase transition}\/ is defined as a nonanalytic point of the canonical free energy $f_N$. The transition is called {\em discontinuous}\/ if the first derivative of $f_N$ is discontinuous, otherwise it is called {\em continuous}.
\end{defn}
In an older but somewhat misleading terminology, discontinuous phase transition are called {\em first-order}\/ phase transitions whereas continuous ones are referred to as {\em second-order}\/ phase transitions.

Computing first and second derivatives of the canonical free energy, one arrives at experimentally measurable quantities like the canonical caloric curve
\begin{equation}
u(\beta)=\frac{\partial}{\partial \beta}\, \beta f_N(\beta)
\end{equation}
or the canonical specific heat
\begin{equation}
c(\beta)=-\beta\frac{\partial}{\partial \beta}\left(\beta^2\frac{\partial f_N(\beta)}{\partial \beta}\right),
\end{equation}
and the effect of the ``order'' of the phase transition is enhanced in the behavior of such derivatives. For example, in the presence of a continuous phase transition, the caloric curve $u(\beta)$ shows a kink at the transition temperature $\beta_\text{t}$, whereas a discontinuity in $u$ is observed in the case of a discontinuous phase transition (hence the name). The liquid--vapor transition of water is an example of a discontinuous phase transition, whereas the transition from ferromagnetic to paramagnetic behavior is a continuous one for many materials.

Note that definition \ref{def:PT} of a phase transition and its order explicitly refers to a particular statistical ensemble, namely the canonical one. This situation seems somewhat unsatisfactory and has been the cause of a longstanding debate. Since, as mentioned in Sec.~\ref{sec:td_functions}, the analyticity properties of thermodynamic functions may depend drastically on the statistical ensembles used, the consequences of this choice of definition are considerable. One can show that the canonical free energy $f_N$ of a system of $N$ degrees of freedom is a smooth function for all finite $N$, and hence no phase transitions occur in finite systems \cite{Griffiths}. This is in contrast to the behavior of {\em microcanonical}\/ thermodynamic functions like the entropy which can have nonanalytic points also in the case of a finite number $N$ of degrees of freedom. We discuss this issue in Sec.~\ref{sec:finite_systems}.

\section{Computation of topological quantities}
\label{sec:models}
Computations of quantities characterizing the topology of the subsets $\MM_v$ as defined in Eq.\ \eqref{eq:Mv} can be found in the literature for several statistical physics models.\footnote{Such model calculations were reported by \textcite{AnAnRuZa:04,Angelani_etal:03,Angelani_etal:05,BaroniPhD,CaCoPe:99,CaCoPe:02,CaPeCo:03,FraPeSpi:00,GaSchiSca:04,GriMo:04,Kastner:04,Kastner:06,KaSchne:06,RiTeiSta:04,RiRiSta:05}.} A numerical computation has been reported by \textcite{FraPeSpi:00}, whereas all other results are analytic, using methods from Morse theory in most of the cases. Morse theory plays an important role not only for the study of model systems, but also for general investigations on the relation between topology and nonanalyticities of thermodynamic functions. For these reasons, a short summary of some basic facts and concepts of this theory is given.

\subsection{Morse theory}
\label{sec:Morse}
Morse theory establishes a link between the two mathematical disciplines of {\em topology}\/ and {\em analysis}. For an introduction to Morse theory, see the textbook by \textcite{Matsumoto} or the classic text by \textcite{Milnor}. For the type of problem we are interested in, Morse theory allows to characterize the topology of the configuration space subsets $\MM_v$ defined in Eq.\ \eqref{eq:Mv} by studying, by means of elementary analysis, the critical points\footnote{Not to be confused with critical points in the theory of phase transitions and critical phenomena.} of the potential $V$.

We consider a smooth (i.\,e., infinitely many times differentiable) function $g:M\to\RR$, mapping some $m$-dimensional manifold $M$ onto the set of real numbers. In the context of our exposition, the role of this general function $g$ will be played by the potential $V$ of a standard Hamiltonian system as defined in Sec.~\ref{sec:standard_H}.
\begin{defn}\label{def:critical_point}
A point $q_{\text{c}}\in M$ is called a {\em critical point}\/ of $g$ if the differential $\dd g\left(q_{\text{c}}\right)$ at $q_{\text{c}}\in M$ vanishes.
\end{defn}
\begin{defn}\label{def:critical_values}
A real number $g_{\text{c}}$ is a {\em critical value} of $g$ if $g\left(q_{\text{c}}\right)=g_{\text{c}}$ for some critical point $q_{\text{c}}$ of $g$.
\end{defn}
With these definitions, we can state a first theorem relating properties of the function $g$ to the topology of the subsets
\begin{equation}
M_t=\left\{q\in M\,\big|\,g(q)\leqslant t\right\}
\end{equation}
[defined analogous to the configuration space subsets $\MM_v$ in Eq.\ \eqref{eq:Mv}].
\begin{thrm}\label{thm:non-critical-neck}
If $g$ has no critical values in the interval $[a,b]$, then $M_a$ and $M_b$ are homeomorphic, i.\,e., there exists a homeomorphism\footnote{A homeomorphism is a continuous bijection between manifolds with continuous inverse.} mapping $M_a$ onto $M_b$.
\end{thrm}
Homeomorphicity is synonymous to {\em topological equivalence}, so $M_a$ and $M_b$ are topologically equivalent, $M_a\sim M_b$, under the above stated conditions: no topology changes take place within the family $\left\{M_t\right\}_{t\in[a,b]}$ upon variation of the parameter $t$ in the interval $[a,b]$.
\begin{proof}[Proof of Theorem~\ref{thm:non-critical-neck}]
see Chap.\ 3.1 of \textcite{Matsumoto}.
\end{proof}

At least in its standard form, Morse theory applies to the class of so-called Morse functions.
\begin{defn}
A critical point $q_\text{c}$ of $g$ is {\em nondegenerate}\/ if the determinant of the Hessian of $g$ at $q_\text{c}$ is nonzero.
\end{defn}
\begin{defn}\label{def:Morse_function}
A function $g:M\to\RR$ is called a {\em Morse function}\/ if every critical point of $g$ is nondegenerate.
\end{defn}
Then Morse theory relates the topology of $M_t$ to the critical points $q_\text{c}$ of $g$ and their indices.
\begin{defn}\label{def:critical_indices}
The {\em index}\/ of a critical point $q_\text{c}$ of $g$ is the number of negative eigenvalues of the Hessian of $g$ at $q_\text{c}$.
\end{defn}
\begin{thrm}\label{thm:handle_decomposition}
If the interval $[a,b]$ contains a single critical value of $g$ with a single critical point $q_\text{c}$, then the topology of $M_b$ differs from the topology of $M_a$ in a way which is determined by the index $i$ of the critical point: $M_b$ is homeomorphic to the manifold obtained from attaching to $M_a$ an $i$-handle, i.\,e., the direct product of an $i$-disk and an $(m-i)$-disk.
\end{thrm}
\begin{proof}
A proof of this theorem, together with a precise definition of ``attaching a handle,'' can be found in Sec.~3.1 of \textcite{Matsumoto}. A generalization to critical values with more than one critical point is straightforward and involves the attachment of more than one handle.
\end{proof}
With Theorems~\ref{thm:non-critical-neck} and \ref{thm:handle_decomposition}, we have transformed the problem of determining the topology of $M_t$ to the problem of determining the critical points and critical indices of the underlying function $g$, which brings us back from topology onto the familiar grounds of analysis.

At least in this standard version, 
the results of Morse theory apply only to the class of Morse functions specified in Definition~\ref{def:Morse_function}. Conceptually, this is an insignificant restriction, since Morse functions on $M$ form an open dense subset of the space of smooth functions on $M$ \cite{Demazure}. This means that, if the potential $V$ of the Hamiltonian system we are interested in is not a Morse function, we can transform it into a Morse function $\bar{V}$ by adding an arbitrarily small perturbation, for example,
\begin{equation}\label{eq:perturbation}
\bar{V}(q)=V(q)+\sum_{i=1}^N h_i q_i
\end{equation}
with some small $h_i\in\RR$ ($i=1,\dotsc,N$). For practical purposes, however, adding a per\-tur\-ba\-tion---and thereby destroying a symmetry present in $V$---may render the explicit computation of critical points and indices much more complicated (or even impossible).  

Theorem~\ref{thm:handle_decomposition} asserts that the topology of the manifolds $M_t$ is characterized by the critical points and their indices, but the ``information'' contained in these quantities---especially in the case of high dimensional manifolds---may be somewhat difficult to handle. A simpler, nonetheless useful characterization of the topology is given by the Euler characteristic, which can be expressed by means of Morse numbers [see any textbook on algebraic topology for a definition of the Euler characteristic for quite general topological spaces, for example, \textcite{Vick}].
\begin{defn}\label{def:Morse_numbers}
The {\em Morse numbers}\/ $\mu_i$ ($i=1,\dotsc,m$) of a function $g$ on an $m$-di\-men\-sion\-al manifold (or $m$-manifold) $M$ are defined as the numbers of critical points of $g$ with index $i$.
\end{defn}
\begin{thrm}\label{thm:Euler-Morse}
The Euler characteristic $\chi$ of an $m$-manifold $M$ can be expressed as the alternating sum of the Morse numbers $\mu_i$ of any Morse function on $M$,
\begin{equation}
\chi=\sum_{i=0}^m (-1)^i \mu_i.
\end{equation}
\end{thrm}
\begin{proof}
See, for example, Chap.~2, Sec.~9.3 of \textcite{Fomenko}.
\end{proof}
An important property of the Euler characteristic $\chi$ is that it is a {\em topological invariant}, i.\,e., different values of $\chi$ for manifolds $M_1$ and $M_2$ imply that $M_1$ and $M_2$ are topologically nonequivalent. Hence monitoring the Euler characteristic of the family $\left\{\MM_v\right\}_{v\in\RR}$ of configuration space subsets of some Hamiltonian system under variation of the parameter $v$, we may get an impression of the way the topology of the $\MM_v$ changes.

\subsection{Model calculation: Mean-field $k$-trigonometric model}
\label{sec:ktrig}
In an explicit calculation, depending on the methods applied (or applicable), different quantities characterizing the topology of the subsets $\MM_v$ may be obtained. Typical examples are as follows:
\begin{enumerate}
\renewcommand{\labelenumi}{(\roman{enumi})}
\item the critical points $q_{\text{c}}$ of $V$ and their indices (Definitions~\ref{def:critical_point} and \ref{def:critical_indices});
\item the critical values $v_{\text{c}}=V(q_{\text{c}})/N$ at which the topology of the $\MM_v$ changes (Definition~\ref{def:critical_values}); and
\item the Morse numbers $\mu_i$ of the $\MM_v$ (Definition~\ref{def:Morse_numbers}), which, by Theorem~\ref{thm:Euler-Morse}, allow one to calculate the Euler characteristic of $\MM_v$.
\end{enumerate}
Such results have been reported in the literature for several model systems. As an illustrating example, we review results for the critical points and for the Euler characteristic of the mean-field $k$\nobreakdash-trigonometric model.

This model is characterized by the potential
\begin{equation}\label{eq:V_k}
V_k(q)=\frac{\Delta}{N^{k-1}}\sum_{i_1,\dotsc,i_k=1}^N \left[1-\cos\left(q_{i_1}+\dotsb+q_{i_k}\right)\right],
\end{equation}
where $\Delta>0$ is some coupling constant and $N$ is the number of degrees of freedom of the system. The position coordinates $q_i\in[0,2\pi)$ ($i=1,\dotsc,N$) are angular variables, so that the configuration space has the shape of an $N$\nobreakdash-dimensional torus. The potential describes a $k$\nobreakdash-body interaction where $k\in\NNN$, and the interaction is of mean-field type, i.\,e., each degree of freedom interacts with each other at equal strength. For this model, a number of thermodynamical as well as topological quantities have been computed by \textcite{Angelani_etal:03,Angelani_etal:05}, the latter ones by making use of Morse theory.\footnote{$V_k$ as given in Eq.\ \eqref{eq:V_k} is not a Morse function, but, in the spirit of Eq.\ \eqref{eq:perturbation}, it can be perturbed into one; see \textcite{Angelani_etal:05}.} Among those results we mention the following:
\begin{enumerate}
\renewcommand{\labelenumi}{(\roman{enumi})}
\item In the thermodynamic limit $N\to\infty$, the mean-field $k$\nobreakdash-trigonometric model does not show a phase transition for $k=1$, whereas it has a phase transition for $k\geqslant2$. The transition is continuous for $k=2$ and discontinuous for all $k\geqslant3$. The potential energy at which the phase transition occurs is $v_{\text{t}}=\Delta$, where the index $\text{t}$ is for {\em transition}.
\item For any finite $N$, the $i$th component of the critical points $q^{m,\ell}_\text{c}$ of the potential $V_k$ is given by
\begin{multline}
\big(q^{m,\ell}_\text{c}\big)_i=
\pi\left\{m_i-\frac{k-1}{k}\left[2\ell+\Heaviside\left(\sum_{j=1}^N m_j-\frac{N}{2}\right)\right]\right\}\\
\pmod {2\pi},
\end{multline}
where $\ell\in\ZZ$, $m=\left(m_1,\dotsc,m_N\right)\in\left\{0,1\right\}^N$, and $\Heaviside$ denotes the Heaviside step function. For our purposes it is important to notice that the number of critical points increases unboundedly with $N$. From the critical points and their indices, the Euler characteristic $\chi\left(\MM_v\right)$ can be computed.
\item In the thermodynamic limit, the critical values
\begin{equation}
v^{m,\ell}_\text{c}=\frac{1}{N}V_k\left(q^{m,\ell}_\text{c}\right)
\end{equation}
become dense on the interval $\left[0,2\Delta\right]$. This leads to a continuously varying (on $\left[0,2\Delta\right]$) limiting distribution
\begin{equation}\label{eq:sigma}
\begin{split}
\sigma(v)&=\lim_{N\to\infty}\frac{1}{N}\ln\left|\chi\left(\MM_v\right)\right|\\
&=-n(v)\ln n(v)-\left[1-n(v)\right]\ln\left[1-n(v)\right]
\end{split}
\end{equation}
of the modulus of the Euler characteristic $\chi\left(\MM_v\right)$ of $\MM_v$, where
\begin{equation}\label{eq:nv}
n(v)=\frac{1}{2}\left[1-\sgn\left(1-\frac{v}{\Delta}\right)\left|1-\frac{v}{\Delta}\right|^{1/k}\right]
\end{equation}
[see \textcite{Angelani_etal:05} for the derivation of this result]. It is a remarkable observation that $\sigma$, which is a purely topological quantity, already signals the absence or presence of a phase transition (see Fig.~\ref{fig:ktrig} for a plot of the graph of $\sigma$): In the case of $k=1$ where the system does not show a phase transition, $\sigma$ is a smooth function. For $k\geqslant2$ there is a nonanalytic point of $\sigma$ at $v=\Delta$, which is precisely the value $v_{\text{t}}$ of the potential energy at which the phase transition occurs.
\begin{figure}[ht]
\center
\psfrag{0.0}{\scriptsize $0.0$}
\psfrag{0.1}{\scriptsize $0.1$}
\psfrag{0.2}{\scriptsize $0.2$}
\psfrag{0.3}{\scriptsize $0.3$}
\psfrag{0.4}{\scriptsize $0.4$}
\psfrag{0.5}{\scriptsize $0.5$}
\psfrag{0.6}{\scriptsize $0.6$}
\psfrag{0.7}{\scriptsize $0.7$}
\psfrag{1.0}{\scriptsize $1.0$}
\psfrag{1.5}{\scriptsize $1.5$}
\psfrag{2.0}{\scriptsize $2.0$}
\psfrag{v/D}{\small $v/\Delta$}
\psfrag{s}{\small $\sigma$}
\psfrag{k=1}{\scriptsize $\negthickspace\negthickspace\negthickspace k=1$}
\psfrag{k=2}{\scriptsize $\negthickspace\negthickspace\negthickspace k=2$}
\psfrag{k=3}{\scriptsize $\negthickspace\negthickspace\negthickspace k=3$}
\psfrag{k=4}{\scriptsize $\negthickspace\negthickspace\negthickspace k=4$}
\includegraphics[width=8.4cm]{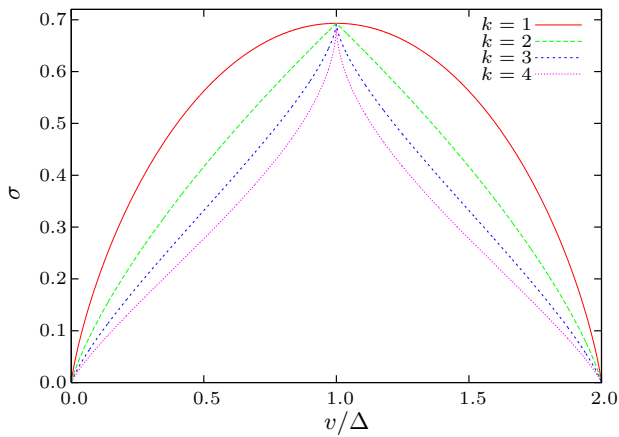}
\caption{\label{fig:ktrig}
(Color online) Logarithmic modulus $\sigma$ of the Euler characteristic of $\MM_v$ as a function of the potential energy $v$ for the mean-field $k$-trigonometric model in the thermodynamic limit [see Eqs.~\eqref{eq:sigma} and \eqref{eq:nv}].}
\end{figure}
\end{enumerate}
In particular this last observation---together with similar findings for other models---may serve as a motivation to study the relation between thermodynamic phase transitions and the topology of the configuration space subsets $\MM_v$.

\subsection{Numerical computation of topological quantities}
\label{sec:numerics}
Only one numerical study of the configuration space topology and its relation to phase transitions has been published to date \cite{FraPeSpi:00}. The key idea of this work is sketched in the following, and the results of an application to the $\varphi^4$ model on the square lattice are summarized. As an outlook, a different numerical approach is sketched for which an application is currently under way. 

\subsubsection{Euler characteristic via Gauss-Bonnet theorem}
A remarkable theorem, found independently by Gauss and Bonnet, connects geometrical and topological quantities of manifolds.
\begin{thrm}
Let $M$ be a compact, orientable Riemannian manifold of even dimension $n$ without boundary. Then its Euler characteristic $\chi(M)$ is given by
\begin{equation}\label{eq:Gauss-Bonnet}
\chi(M) = \frac{2}{\left|{\mathbbm S}^n\right|}\int_M \dd\sigma K(\sigma),
\end{equation}
where $\left|{\mathbbm S}^n\right|$ denotes the volume of an $n$-dimensional unit sphere and $K$ is the Gauss-Kronecker curvature.
\end{thrm}
The Gauss-Kronecker curvature is a measure of how much the normal vector of the manifold at a point $\sigma$ changes upon infinitesimal variation of $\sigma$.\footnote{For a precise definition see any introductory text on differential geometry, e.\,g., Chapter 3F of \textcite{Kuehnel}.} This theorem is remarkable, and useful for our purposes, since it expresses {\em global}\/ topological properties of a manifold in terms of {\em local}\/ geometrical ones. The locality of the Gauss-Kronecker curvature $K$ allows to estimate (apart from a prefactor) the integral in \eqref{eq:Gauss-Bonnet} by implementing some dynamics on $M$ and probing $K$ in the course of the dynamical evolution.

This method has been proposed and implemented by \textcite{FraPeSpi:00} for the numerical computation of the Euler characteristic of the configuration space subsets $\Sigma_v$ of the $\varphi^4$ model in two dimensions on the square lattice with nearest-neighbor interactions. This model is characterized by the potential\footnote{This definition differs slightly from the one used by \textcite{FraPeSpi:00}, but both models can be mapped onto each other by a suitable rescaling of $J$, $V$, and $q$.}
\begin{equation}\label{eq:phi4nn}
V_\varphi^{\text{nn}}(q)=\sum_{i=1}^N \left(-\frac{1}{2}q_i^2 + \frac{1}{4}q_i^4\right)-J\sum_{\langle i,j\rangle} q_i q_j,
\end{equation}
where $J\in\RR$ is a coupling constant and the angular brackets $\langle i,j\rangle$ denote a summation over all pairs of nearest neighbors on the square lattice. The first term of the potential is an on-site potential with the shape of a double well, and the second term describes a pair interaction between each degree of freedom $q_i\in\RR$ and its four nearest neighbors on the lattice. The superscript of $V_\varphi^{\text{nn}}$ serves to distinguish the $\varphi^4$ model with nearest-neighbor interactions from a similar one with long-range interactions introduced in Sec.~\ref{sec:long-range}. For positive values $J>0$ of the coupling constant, a parallel orientation of neighboring degrees of freedom is energetically favorable, and the model is known to show a phase transition from a ferromagnetic phase at low temperatures to a paramagnetic phase at high temperatures in the thermodynamic limit.

\begin{figure}[htb]
\center
\includegraphics[angle=270,width=8.4cm]{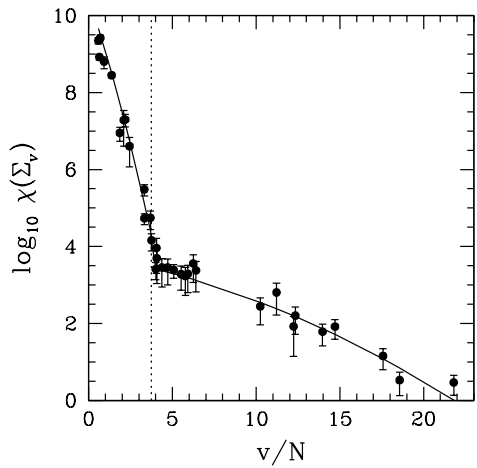}
\caption{\label{fig:Euler}
Logarithmic modulus $\sigma$ of the Euler characteristic of $\Sigma_v$ as a function of the potential energy per degree of freedom for the nearest-neighbor $\varphi^4$ model on a $7\times 7$ square lattice. The dotted vertical line marks the potential energy at which a phase transition takes place in the thermodynamic limit. The solid line serves as a guide to the eye. From \textcite{FraPeSpi:00}.}
\end{figure}

Implementing a Monte Carlo dynamics on $\Sigma_v$, \textcite{FraPeSpi:00} computed numerical estimators for the relative variation of the Euler characteristic $\chi(\Sigma_v)$ of the nearest-neighbor $\varphi^4$ model (see their publication for more details on the implementation). Their result for a system of $7\times7$ lattice sites is reproduced in Fig.~\ref{fig:Euler}.%
The data suggest that a kink is present in $\chi(\Sigma_v)$ at a value of the potential energy per particle $v$ which is very close to the value of $v$ at which a phase transition takes place in the thermodynamic limit.

\subsubsection{Critical points of the potential}
Morse theory, as introduced in Sec.~\ref{sec:Morse}, suggests another, straightforward way for the numerical estimation of topological quantities: Determining numerically the critical points and their indices of a potential $v$, one can calculate the Euler characteristic of the configuration space subsets $\MM_v$ by means of Theorem~\ref{thm:Euler-Morse}. Since the number of critical points in generic models is assumed to grow exponentially with the number $N$ of degrees of freedom, this task is computationally very hard, and for $N$ of order $10^2$ or larger one can expect to find at best a small fraction of the existing critical points.

Numerical methods for the computation of critical points of high-dimensional functions have been employed and tested in the study of the properties of potential energy landscapes of glassy systems, clusters, and others \cite{Gri_etal:02,Wales}. An application of these methods to the study of configuration space topology and phase transitions is currently under way.

\section{Nonanalyticities in finite systems}
\label{sec:finite_systems}
Before turning to the study of the relation between configuration space topology and phase transitions, i.\,e., the behavior of thermodynamic functions in the thermodynamic limit, we discuss the somewhat simpler case of finite systems. The central question addressed in this section is:

{\em What are the differentiability properties of the thermodynamic functions $s_N$ and $f_N$ for finite numbers $N$ of degrees of freedom?}

For the configurational canonical free energy the answer is well known: $f_N$ is a smooth function on its entire domain for all finite values of $N$ \cite{Griffiths}. For technical reasons, the microcanonical ensemble has been used only sporadically for explicit calculations in statistical physics, and little attention has been paid to the differentiability properties of the configurational microcanonical entropy. In this section we investigate the relation between nonanalytic points of $s_N$ and critical points of $V$ (which, in turn, are related to topology changes of the configuration space subsets $\MM_v$):

{\em What is the effect of a critical point $q_{\text{c}}$ of the potential $V$ on the differentiability of $s_N$ at the corresponding critical value $v_{\text{c}}=V\left(q_{\text{c}}\right)/N$?}

Recently, exact model calculations have been performed which demonstrate that the microcanonical entropy $\bar{s}_N$ or the configurational microcanonical entropy $s_N$ of classical statistical physics models can have nonanalytic points for finite $N$ \cite{KaSchne:06,DunHil1:06,DunHil2:06,CaKa:06}. At least in the case of smooth potentials $V$ as considered by \textcite{KaSchne:06} and \textcite{CaKa:06} where the mean-field spherical model was studied, nonanalyticities of $s_N$ show up precisely at the critical values $v_{\text{c}}$ of $V$.

The occurrence of nonanalyticities in $s_N$---though a surprise even to many researchers working in the field---can be anticipated from the discussion of a simple one-dimensional example. Considering a $\varphi^4$ potential on the real line,
\begin{equation}
V(q)=-\tfrac{1}{2}q^2 + \tfrac{1}{4}q^4,
\end{equation}
the calculation of the configurational microcanonical entropy yields
\begin{multline}
s_1(v)=\ln\Bigg[\frac{2\Heaviside(v+\frac{1}{4})}{\sqrt{(1+4v)}}\\
\times\left(\frac{1}{\sqrt{1+\sqrt{1+4v}}}+\frac{\Heaviside(-v)}{\sqrt{1-\sqrt{1+4v}}}\right)\Bigg].
\end{multline}
This function has a nonanalytic point at argument zero, which is identical to the critical value of $V$ corresponding to the critical point $q=0$. From this one-dimensional example, it appears plausible that nonanalyticities show up also at critical values of higher dimensional potentials. This reasoning is confirmed by the following theorem by \textcite{KaSchneSchrei:07}.
\begin{thrm}\label{thm:finite}
Let $V:G\to\RR$ be a Morse function with a single critical point $q_\text{c}$ of index $i$ in an open region $G\subset\RR^N$. Without loss of generality, we assume $V(q_\text{c})=0$. Then there exists a polynomial $P$ of degree less than $N/2$ such that at $v=0$ the density of states $\Omega_N=\exp(Ns_N)$ can be written in the form
\begin{equation}\label{eq:Omega_sep}
\Omega_N(v)=P(v)+\frac{h_{N,i}(v)}{\sqrt{\left|\det\left[{\mathfrak H}_{V}(q_\text{c})\right]\right|}}+\smallo(v^{N/2-\epsilon})
\end{equation}
for any $\epsilon>0$. Here ${\mathfrak H}_{V}$ is the Hessian of $V$, $\smallo$ denotes Landau's little-o symbol for asymptotic negligibility, and
\begin{multline}\label{eq:h_Ni}
h_{N,i}(v)=\frac{(N\pi)^{N/2}}{\Gamma(N/2)}\\
\times\begin{cases}
(-1)^{(N-i)/2} (-v)^{(N-2)/2} \Heaviside(-v) & \text{for $N,i$ odd},\\
(-1)^{i/2} \,v^{(N-2)/2} \Heaviside(v) & \text{for $i$ even},\\
(-1)^{(i+1)/2} \,v^{(N-2)/2}\,\pi^{-1}\ln|v| & \text{for $N$ even, $i$ odd},\\
\end{cases}
\end{multline}
is a universal function.
\end{thrm}
\begin{proof}
For a proof of this theorem, the density of states is calculated separately below and above the critical value $v=0$. By complex continuation it is possible to subtract both contributions and to evaluate the leading order of the difference. A detailed proof will be published elsewhere. A related, but weaker result has been announced by \textcite{Spinelli}.
\end{proof}
We see from this theorem that, for all finite system sizes, a critical point of $V$ produces a nonanalyticity in the entropy $s_N$ at the corresponding critical value. The theorem refers to a {\em single}\/ critical point, but this condition can be readily released by considering the union of regions around several critical points.

Remarkably, the type of nonanalyticity described by $h_{N,i}$ in Eq.\ \eqref{eq:h_Ni} does not depend on the precise value of the index $i$ of the critical point, but only on whether $N$ and $i$ are odd or even. One can verify that, independently of the three cases in Eq.\ \eqref{eq:h_Ni}, $h_{N,i}$, and therefore also the entropy $s_N$, is precisely $\left\lfloor(N-3)/2\right\rfloor$ times continuously differentiable. This result is in agreement with the nonanalytic behavior of the exact solution for the entropy of the mean-field spherical model as reported by \textcite{KaSchne:06}. In other words, $s_N$ becomes ``smoother'' with increasing number $N$ of degrees of freedom, and already for moderate $N$ it supposedly will be impossible to observe such a finite-system nonanalyticity from noisy experimental or numerical data.

\section{Phase transitions and configuration space topology}
\label{sec:theorems}
We have seen in the previous section that, for finite systems, critical values of the potential $V$ (and hence topology changes of the configuration space subsets $\MM_v$) are directly related to the occurrence of nonanalytic points of the configurational microcanonical entropy $s_N$. For infinite systems, such a straightforward correspondence cannot be expected since, according to Theorem~\ref{thm:finite}, the order of differentiability of $s_N$ in the presence of a finite number of critical points of $V$ diverges when $N\to\infty$. Furthermore, from the results on the mean-field $k$\nobreakdash-trigonometric model reviewed in Sec.~\ref{sec:ktrig}, it is obvious that a relation between topology changes and phase transitions for the infinite system case has to be subtler: The number of critical values $v_{\text{c}}$ of the potential energy where topology changes of $\MM_v$ occur increases unboundedly with the number $N$ of degrees of freedom of the system and becomes dense on the interval $[0,2\Delta]$ in the limit $N\to\infty$. Since a phase transition in this model occurs only at {\em one}\/ particular value of the potential energy $v=v_{\text{t}}$, it is clear that there cannot be a one-to-one relation between phase transitions in infinite systems and topology changes. Nonetheless, the existence of some sort of relation is suggested by the singular behavior of the topological invariant $\sigma$ at $v_{\text{t}}$ as plotted in Fig.~\ref{fig:ktrig}. Similar results, i.\,e., a continuum of topology changes of the $\MM_v$ and a nonanalyticity in a quantity characterizing the topology in the thermodynamic limit, were found for the mean-field $XY$ model \cite{CaCoPe:99,CaCoPe:02,CaPeCo:03}. Numerical studies also indicate that the same features are present in the $\varphi^4$ model on a square lattice with nearest-neighbor interactions \cite{FraPeSpi:00}.

\subsection{Conjectures}
\label{sec:conjectures}
From the results of such model calculations, a general, but somewhat unspecified ``relation'' between topology changes of the $\MM_v$ and phase transitions was conjectured by \textcite{CaCaClePe:97}. Later on, this conjecture became known as the {\em topological hypothesis}, and its formulation gave leeway for different interpretations of its content.

The results on the mean-field $k$\nobreakdash-trigonometric model, the mean-field $XY$ model, and the square lattice $\varphi^4$ mod\-el cited above might suggest the following formulation.
\begin{conjecture}\label{con:strong}
If, in the thermodynamic limit, the logarithmic density of the Euler characteristic $\chi$ of $\MM_v$ has a nonanalytic point at $v=v_{\text{t}}$, a phase transition takes place at the potential energy $v_{\text{t}}$. 
\end{conjecture}
A weaker version of such a hypothesis might be formulated as follows.
\begin{conjecture}\label{con:weak}
A topology change within the family $\left\{\MM_v\right\}_{v\in\RR}$ at $v=v_{\text{t}}$ is a {\em necessary}\/ condition for a phase transition to take place at $v_{\text{t}}$. 
\end{conjecture}
We will see shortly that neither of these conjectures holds true for {\em arbitrary}\/ statistical systems, and counterexamples disproving their general validity will be given in Sec.~\ref{sec:limitations}. Nonetheless the conjectures are of interest as they prove to be correct at least for certain (large and relevant) {\em classes}\/ of systems. In fact, a theorem presented in the next section asserts that Conjecture~\ref{con:weak} is correct for a class of models with short-range interactions.

\subsection{Franzosi-Pettini theorem}
\label{sec:theorem}
In 2004, Franzosi and Pettini announced the proof of a theorem asserting that Conjecture~\ref{con:weak} is correct for a class of short-range models \cite{FraPe:04}. In the present section we sketch this result, and refer the reader to \textcite{FraPeSpi} and \textcite{FraPe} for details. We start with the definitions necessary to specify the class of models for which the theorem holds.

\begin{figure}[ht]
\center
\psfrag{V}{\scriptsize $V$}
\psfrag{q}{\scriptsize $q$}
\includegraphics[width=4cm]{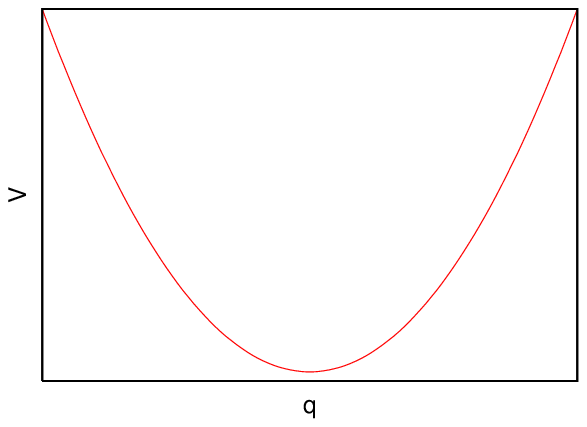}
\hfill
\includegraphics[width=4cm]{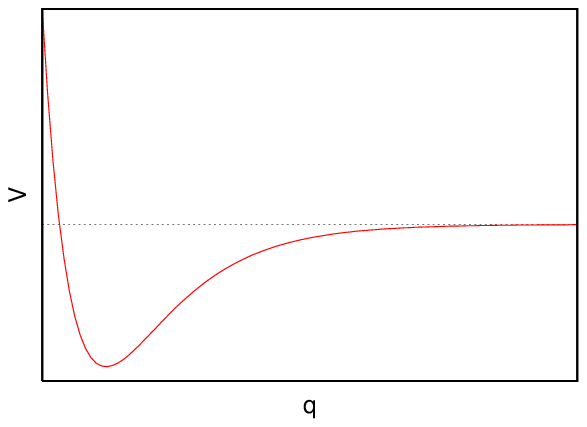}
\caption{\label{fig:confining}
(Color online) One-dimensional illustrations of confining and non-confining potentials. Left: The parabola $V(q)=q^2$ is confining, as $\lim\limits_{q\to\pm\infty}V(q)=\infty$. Right: The Morse potential $V(q)=\left(1-\ee^{-q}\right)^2$ is nonconfining, as $\lim\limits_{q\to\infty}V(q)<\infty$.}
\end{figure}

As throughout the paper, we consider systems of $N$ degrees of freedom described by a Hamiltonian of standard form \eqref{eq:H_standard}. We assume the potential $V$, defined on the continuous configuration space $\Gamma_N$, to be subject to a number of conditions, whose definitions are given in the following. 
\begin{defn}\label{def:standardform}
The potential $V$ is of {\em standard form}
\begin{equation}
V(q)=\sum_{i=1}^N\phi\left(q_i\right) + \sum_{i,j=1}^N c_{ij}\psi\left(\left\lVert q_i-q_j\right\rVert\right)
\end{equation}
if it consists of an on-site potential $\phi$ and a pair potential $\psi$, the latter depending only on the Euclidean distance $\lVert\cdot\rVert$ of the degrees of freedom. For a lattice system the coefficients $c_{ij}$ determine the coupling strength between the degrees of freedom on different lattice points. For a fluid (nonlattice) system we typically have $c_{ij}=1-\delta_{i,j}$, where $\delta_{i,j}$ is the Kronecker symbol.
\end{defn}
\begin{defn}\label{def:shortrange}
The potential $V$ is of {\em short range}\/ if
\begin{description}
\item {\em on a lattice:} the coefficients $c_{ij}$ are nonzero only for $i,j$ from a finite neighborhood on the lattice;
\item {\em in a fluid:} for large $x$, $\left\lvert\psi\left(x\right)\right\rvert$ decreases faster than $x^{-d}$, where $d$ is the spatial dimension of the system.
\end{description}
\end{defn}
\begin{defn}
The potential $V$ is {\em stable}\/ if, for any $N\in\NNN$ and all $q\in\Gamma_N$, there exists a constant $B\geqslant0$ such that
\begin{equation}
V(q)\geqslant-NB.
\end{equation}
\end{defn}
\begin{defn}\label{def:confining}
The potential $V$ is said to be {\em confining}\/ if $\MM_v$ as defined in Eq.\ \eqref{eq:Mv} is a compact set for all finite values of $v$.
\end{defn}
Physically speaking, a particle in a confining potential cannot escape to infinity at finite energy. Simple one-dimensional examples of confining and nonconfining potentials are plotted in Fig.~\ref{fig:confining}.

It is well known that for potentials of standard form which are stable, confining, and of short range, the thermodynamic limit of thermodynamic functions like the configurational microcanonical entropy \eqref{eq:s_conf} or the configurational canonical free energy \eqref{eq:f_conf} exists, i.\,e.,
\begin{equation}\label{eq:sinf}
s_\infty(v)=\lim_{N\to\infty}s_N(v)<\infty
\end{equation}
and
\begin{equation}\label{eq:finf}
f_\infty(\beta)=\lim_{N\to\infty}f_N(\beta)<\infty
\end{equation}
[see \textcite{Ruelle} for a proof of these and related results for even larger classes of systems]. For this class of systems, Franzosi and Pettini proved a theorem which can be phrased as follows.
\begin{thrm}\label{thrm:1}
Let $V$ be a potential of standard form which is smooth, stable, confining, bounded below, and of short range. If there exists an interval $\left[v_1,v_2\right]$ such that, for any $N$ larger than some constant $N_0$, the $\left\{\MM_v\right\}_{v\in\left[v_1,v_2\right]}$ do not change topology, then in the thermodynamic limit the canonical configurational free energy $f_N$ is at least two times continuously differentiable in the interval $\left(\beta\left(v_1\right),\beta\left(v_2\right)\right)$, where $\beta\left(v_1\right)$ and $\beta\left(v_2\right)$ are the values of the inverse temperature corresponding to the potential energies $v_1$ and $v_2$, respectively.
\end{thrm}
\begin{proof}
The  proof makes use of the fact that, in an interval free of critical values of $V/N$, the configurational entropy $s_N$ is smooth for all finite $N$ (which is a result known from Morse theory). Then it is shown that at least the first four derivatives of $s_N$ are uniformly bounded above in $N$. This implies that $s_\infty$ is three times continuously differentiable, and therefore $f_\infty$ is two times continuously differentiable in the corresponding interval of inverse temperatures. A detailed proof can be found in \textcite{FraPeSpi}.
\end{proof}
Although this theorem does not exclude the possibility of a discontinuity in some higher derivative of $f_\infty$, one would assume that the extension of the theorem to higher derivatives, though laborious, is straightforward. Such a generalization to arbitrary derivatives of $f_\infty$ would assert the correctness of Conjecture~\ref{con:weak} for the class of potentials covered by the theorem.

Theorem~\ref{thrm:1} rigorously establishes a connection between phase transitions and topology changes of the $\MM_v$ for a certain class of systems, which from a conceptual point of view is a remarkable result. For potentials like that of a $\beta$-Fermi-Pasta-Ulam chain, having no further critical points apart from their overall minimum, Theorem~\ref{thrm:1} allows one to rigorously exclude the occurrence of a phase transition. Possibly even more important, for systems like the $\varphi^4$ model in Eq.\ \eqref{eq:phi4nn} one may deduce an upper bound on the transition potential energy $v_\text{t}$ from the presence of an upper bound on the critical values of $V$.

For the typical models of interest, the critical values $v_{\text{c}}$ of the potential become dense in the thermodynamic limit, in some cases on an interval, but often on the entire codomain of $V/N$ (like in the case of the mean-field $k$-trigonometric model discussed in Sec.~\ref{sec:ktrig}). In that case, no interval $\left[v_1,v_2\right]$, free of critical points in the sense specified in Theorem~\ref{thrm:1}, exists, and no conclusions can be drawn from the theorem. An attempt to derive a stronger result which sheds light on the relation between phase transitions and topology changes even in the presence of (possibly very many) critical points has been made by \textcite{FraPe}, and in a modified form by \textcite{KaSchneSchrei:07}, and the hope is to understand from these studies how topology might ``act'' to produce a nonanalyticity in some thermodynamic function in the thermodynamic limit. 

\subsection{Models not covered by Theorem~\ref{thrm:1}}
Theorem~\ref{thrm:1} establishes a necessary relation between phase transitions and configuration space topology for a certain class of systems with short-range interactions and confining potentials. Of course, this result does not exclude the possibility that a phase transition in a long-range system or in a system with a nonconfining potential may nonetheless {\em be}\/ related to a topology change in configuration space. In fact, all models for which analytic calculations of the topology of the $\MM_v$ have been reported so far do {\em not}\/ meet the assumptions of Theorem~\ref{thrm:1}, being either long range, nonconfining, nonsmooth, or even some combination of these properties. Results from model calculations reported in the literature are summarized in Table~\ref{tab:models}, and we notice that for several of these models which do not comply with the assumptions of Theorem~\ref{thrm:1}, a relation between phase transitions and topology is nonetheless observed, even in the sense of the stronger Conjecture~\ref{con:strong}. Some other models, however, disprove the general validity of Conjectures~\ref{con:strong} and \ref{con:weak}, and we look at these cases in more detail in Sec.~\ref{sec:limitations}.%
\renewcommand{\arraystretch}{1.1}
\setlength{\tabcolsep}{1.3mm}
\ctable[caption={(Color online) Results from calculations of the topology of $\MM_v$ for some models (for definitions of the models, see the references indicated). The abbreviation {\em m.-f.} is for mean-field interactions and {\em n.\,n.} is for nearest-neighbor interactions. The columns list (from left to right) whether the respective model has a phase transition, whether the results on the topology of $\MM_v$ were obtained analytically or numerically, whether the potential of the system is short range, smooth, and confining, and whether the results are in accordance with Conjectures~\ref{con:strong} and \ref{con:weak}.}, label=tab:models,pos=ht,star]
{|l||*{5}{c|}|c|c||c|}
{
\tnote{Although the pair potential of the nearest-neighbor spherical model appears to be short range, an effective long rangedness is induced by the spherical constraint.}
\tnote[b]{As far as discernable from numerical data.}
}
{
\hline
& phase &anal./ & short- & & & Conj. & Conj. &\\
\raisebox{1.5ex}[-1.5ex]{model} & trans. & num. & range & \raisebox{1.5ex}[-1.5ex]{smooth} & \raisebox{1.5ex}[-1.5ex]{conf.} & \ref{con:strong} & \ref{con:weak} & \raisebox{1.5ex}[-1.5ex]{references}\\\hline\hline
m.-f. $XY$ & {\color{OliveGreen}yes} & {\color{OliveGreen}anal.} & {\color{BrickRed}no} & {\color{OliveGreen}yes} & {\color{OliveGreen}yes} & {\color{OliveGreen}yes} & {\color{OliveGreen}yes} & \parbox{6.5cm}{{\vspace{1mm}\small \cite{CaCoPe:99,CaCoPe:02,CaPeCo:03}\vspace{.6mm}}} \\\hline
m.-f. $k$-trigon. & {\color{OliveGreen}yes} & {\color{OliveGreen}anal.} & {\color{BrickRed}no} & {\color{OliveGreen}yes} & {\color{OliveGreen}yes} & {\color{OliveGreen}yes} & {\color{OliveGreen}yes} & \parbox{6.5cm}{{\vspace{1mm}\small \cite{Angelani_etal:03,Angelani_etal:05}\vspace{0.6mm}}} \\\hline
m.-f. spherical & {\color{OliveGreen}yes} & {\color{OliveGreen}anal.} & {\color{BrickRed}no} & {\color{OliveGreen}yes} & {\color{OliveGreen}yes} & {\color{OliveGreen}yes} & {\color{OliveGreen}yes} & \parbox{6.5cm}{{\vspace{1mm}\small \cite{RiTeiSta:04,Kastner:06,KaSchne:06}\vspace{0.6mm}}} \\\hline
m.-f. $\varphi^4$ & {\color{OliveGreen}yes} & {\color{OliveGreen}anal.} & {\color{BrickRed}no} & {\color{OliveGreen}yes} & {\color{OliveGreen}yes} & {\color{BrickRed}no} & {\color{BrickRed}no} & \parbox{6.5cm}{{\vspace{1mm}\small \cite{BaroniPhD,GaSchiSca:04,AnAnRuZa:04,HaKa:05}\vspace{0.6mm}}} \\\hline
n.\,n. spherical & {\color{OliveGreen}yes} & {\color{OliveGreen}anal.} & {\color{BrickRed}no}\tmark & {\color{OliveGreen}yes} & {\color{OliveGreen}yes} & {\color{BrickRed}no} & {\color{OliveGreen}yes} & \parbox{6.5cm}{{\vspace{1mm}\small \cite{RiRiSta:05}\vspace{0.6mm}}} \\\hline
Peyrard-Bishop & {\color{OliveGreen}yes} & {\color{OliveGreen}anal.} & {\color{OliveGreen}yes} & {\color{OliveGreen}yes} & {\color{BrickRed}no} & {\color{BrickRed}no} & {\color{BrickRed}no} & \parbox{6.5cm}{{\vspace{1mm}\small \cite{GriMo:04,AnRuZa:05}\vspace{0.6mm}}} \\\hline
Burkhardt & {\color{OliveGreen}yes} & {\color{OliveGreen}anal.} & {\color{OliveGreen}yes} & {\color{BrickRed}no} & {\color{BrickRed}no} & {\color{BrickRed}no} & {\color{BrickRed}no} & \parbox{6.5cm}{{\vspace{1mm}\small \cite{Kastner:04,AnRuZa:05}\vspace{0.6mm}}} \\\hline
$2d$ n.\,n. $\varphi^4$ & {\color{OliveGreen}yes} & {\color{BrickRed}num.} & {\color{OliveGreen}yes} & {\color{OliveGreen}yes} & {\color{OliveGreen}yes} & {\color{OliveGreen}(yes)}\tmark[b] & {\color{OliveGreen}yes} & \parbox{6.5cm}{{\vspace{1mm}\small \cite{FraPeSpi:00}\vspace{0.6mm}}} \\\hline
$1d$ n.\,n. $XY$ & {\color{BrickRed}no} & {\color{OliveGreen}anal.} & {\color{OliveGreen}yes} & {\color{OliveGreen}yes} & {\color{OliveGreen}yes} & --- & --- & \parbox{6.5cm}{{\vspace{1mm}\small \cite{CaPeCo:03}\vspace{0.6mm}}} \\\hline
}

The above observations, together with Theorem~\ref{thrm:1}, can be interpreted in the following way: A change of the topology of the $\MM_v$ under variation of the parameter $v$ is one possible ``mechanism'' which can lead to a nonanalyticity in a thermodynamic function. However, the fact that some of the models listed in Table~\ref{tab:models} are not in accordance with Conjectures~\ref{con:strong} and \ref{con:weak} indicates that a topology change is not the only such mechanism, and we get to know a second nonanalyticity generating mechanism in Sec.~\ref{sec:sing_max}. Theorem~\ref{thrm:1} then suggests that, for the---from a physical point of view very important---class of systems fulfilling its assumptions, a topology change is the {\em only}\/ mechanism available to cause a nonanalyticity.

\section{Limitations of the relation between phase transitions and configuration space topology}
\label{sec:limitations}
We have seen in Sec.~\ref{sec:theorem} that, for a certain class of systems, a topology change in the configuration space subsets $\MM_v$ is necessary for a phase transition to take place at the corresponding energy or temperature. For some models however, as indicated in Table~\ref{tab:models}, a phase transition is not necessarily accompanied by a topology change. These results immediately suggest the following question:

{\em Which of the restrictions on the class of systems for which Theorem~\ref{thrm:1} holds are mere technicalities which could be relaxed by more refined methods of proof, and which ones are really essential?}

In the present section we discuss two types of systems which disprove the general validity of Conjectures~\ref{con:strong} and \ref{con:weak}, thereby showing that at least the short-rangedness and the confining property imposed on the potential $V$ in Theorem~\ref{thrm:1} cannot be relaxed.

\subsection{Long-range interactions}
\label{sec:long-range}
Systems with long-range interactions are often neglected in standard treatises on statistical mechanics. In many cases they are even outside the scope of traditional thermodynamics because, in contrast to the case of short-range interactions (Definition~\ref{def:shortrange}), the thermodynamic limit of thermodynamic functions as in Eqs.\ \eqref{eq:sinf} and \eqref{eq:finf} does not necessarily exist. Furthermore, the {\em stability}\/ of thermodynamic functions, manifest in their convexity properties, is no longer guaranteed [see \textcite{Dauxois_etal} for a review of the dynamics and thermodynamics of systems with long-range interactions]. This limitation of traditional thermodynamics is remarkable in regard to the importance of long-range interactions in physics, most notably in gravitation and in electrodynamics (at least in the absence of screening effects). In this section, we give an example of a long-range system whose phase transition is not accompanied by a topology change in configuration space. In Sec.\ \ref{sec:sing_max} we argue that it is precisely due to the above-mentioned convexity properties of thermodynamic functions that phase transitions in long-range systems are not necessarily related to topology changes.

The example which we discuss is the mean-field $\varphi^4$ model, defined by the potential
\begin{equation}\label{eq:potphi4}
V_\varphi^{\text{mf}}(q)=\sum_{i=1}^N \left(-\frac{1}{2}q_i^2 + \frac{1}{4}q_i^4\right)-\frac{J}{2N}\left(\sum_{i=1}^N q_i\right)^{\!\!\!2}
\end{equation}
with coupling constant $J>0$. The first term of the potential is an on-site potential with the shape of a double well. The second term of $V_\varphi^{\text{mf}}$ describes a pair interaction of mean-field type, i.\,e., each degree of freedom $q_i\in\RR$ is coupled to each other at equal strength, which is an extreme case of long-range interactions. The choice of an $N$-dependent coupling strength $-J/(2N)$, though questionable from a physical point of view, guarantees the existence of the thermodynamic functions in the limit $N\to\infty$.

This model is exactly solvable in the sense that, in the thermodynamic limit, thermodynamic functions like the configurational microcanonical entropy \cite{HaKa:05,HaKa:06,CamRuf:06} or the configurational canonical free energy \cite{OvOn:90,DauxLeRu:03} can be expressed in terms of a maximization and a single integration. A continuous phase transition is found to take place in the mean-field $\varphi^4$ model, and from the configurational microcanonical entropy an implicit expression for the potential energy $v_{\text{t}}$ at which the transition occurs was derived by \textcite{HaKa:06}. For large values of the coupling constant $J$, an expansion of this implicit expression yields
\begin{equation}\label{eq:vc_Jlarge}
v_{\text{t}}(J)=a J^2-\left(2a-\tfrac{1}{4}\right)J+\Order(1),
\end{equation}
with $a=\Gamma(3/4)/\Gamma(1/4)$. For our purposes it is important to note that $v_{\text{t}}$ increases unboundedly with $J$, i.\,e., the transition energy can be made arbitrarily large by increasing the coupling constant $J$.

A study of the topology of the configuration space subsets $\MM_v$ of the mean-field $\varphi^4$ model has first been reported by \textcite{BaroniPhD} and later, independently and by different methods, by \textcite{GaSchiSca:04}. Looking for critical points $q_{\text{c}}$ satisfying $\dd V_\varphi^{\text{mf}}(q_{\text{c}})=0$, one can show that the corresponding critical values
\begin{equation}\label{eq:vc_phi4}
v_{\text{c}}=\frac{V_\varphi^{\text{mf}}(q_{\text{c}})}{N}\leqslant0
\end{equation}
are nonpositive for all critical points $q_{\text{c}}$ of the potential $V_\varphi^{\text{mf}}$ [see Appendix E of \textcite{BaroniPhD} for a proof].

Confronting the results \eqref{eq:vc_Jlarge} and \eqref{eq:vc_phi4}, i.\,e., the unbounded growth of the potential energy $v_{\text{t}}$ of the phase transition with the boundedness from above of the critical values $v_{\text{c}}$ of $V_\varphi^{\text{mf}}$, one immediately arrives at the conclusion that, for the mean-field $\varphi^4$ model, the phase transition is not accompanied by a topology change in configuration space in general (i.\,e., not for arbitrary coupling constants $J$). Hence neither conjecture \ref{con:weak} nor conjecture \ref{con:strong} on the relation between configuration space topology and phase transitions hold true for this model. The mean-field $\varphi^4$ model fulfills, apart from the short rangedness, all requirements of Theorem~\ref{thrm:1}. Hence we conclude that the assumption in Theorem~\ref{thrm:1} of the potential being short ranged cannot be relaxed in general.

\subsection{Nonconfining potentials}
\label{sec:nonconfining}
Particles in a confining potential (Definition~\ref{def:confining}) are restricted to a bounded subset of the configuration space for any finite value of the energy. Modeling gases or fluids by potentials of standard form (Definition~\ref{def:standardform}), the pair interaction between particles is typically assumed to be nonconfining, but the addition of an on-site potential modeling a container renders the overall potential confining. Nonconfining potentials are of physical interest for modeling fluctuations of interfaces by means of so-called solid-on-solid models \cite{Abraham}. The occurrence of a localization-delocalization transition in these models is a consequence of the nonconfining property of the potential. In this section, we discuss a class of solid-on-solid models with nonconfining potentials, for which we find phase transitions not accompanied by topology changes in configuration space.

The solid-on-solid models we consider are one-di\-men\-sion\-al lattice models characterized by a potential of the form
\begin{equation}\label{eq:SOS}
V_\text{\tiny SOS}(q)=J\sum_{i=1}^{N-1} \left|q_{i+1}-q_i\right|^n + \sum_{i=1}^N U\left(q_i\right)
\end{equation}
with some coupling constant $J>0$. For the modeling of physical systems, the parameter $n$ typically has values $n=1$ or $2$. The position coordinates $q_i$ can take on values from the positive half-line, so that the configuration space of the model is $\Gamma_N=\left(\RR^+\right)^N$. The pair interaction, in contrast to the example discussed in Sec.~\ref{sec:long-range}, is of short range, being restricted to nearest neighbors on the lattice. The on-site potential $U$ is a real-valued function with a single local minimum somewhere on its domain $\RR^+$, and it approaches a finite value in the limit of large arguments,
\begin{equation}\label{eq:pinning}
\lim_{x\to\infty}U(x)<\infty
\end{equation}
(like the nonconfining one-dimensional potential in the right plot of Fig.~\ref{fig:confining}). As a consequence of Eq.\ \eqref{eq:pinning}, we have that the potential $V_\text{\tiny SOS}$ in Eq.\ \eqref{eq:SOS} is also nonconfining.

By means of a transfer operator technique \cite{KraWan:41}, the thermodynamic limit value of the configurational canonical free energy $f_\infty$ of the solid-on-solid model \eqref{eq:SOS} can be written as
\begin{equation}\label{eq:EV_max}
-\beta f_\infty\left(\beta\right)=\max_{i}\left[\ln\lambda_i\left(\beta\right)\right],
\end{equation}
where $\lambda_i$ are the eigenvalues of the (well-behaved) solutions of the eigenvalue problem
\begin{multline}\label{eq:transfer}
\int_0^\infty \dd y\,\exp\left\{-\tfrac{1}{2}\beta U(x)-J\beta\left|x-y\right|^n-\tfrac{1}{2}\beta U(y)\right\}\Phi\left(y\right)\\
=\lambda\Phi(x).
\end{multline}
Equation~\eqref{eq:transfer} can be transformed into a Schr\"odinger-type eigenvalue problem. This mapping is exact for the case $n=1$ \cite{Burkhardt:81}, whereas it involves a gradient-expansion approximation for the case $n=2$ [see \textcite{Theodorakopoulos:03} for details]. For the Schr\"odinger-type equation thus obtained, one can argue that, depending on the values of $J$ and $\beta$, bound state solutions may or may not exist, and it is this changeover which corresponds to the occurrence of a phase transition in the statistical mechanical model. For simple box-type potentials $U$, the eigenvalue problem can be solved analytically \cite{Burkhardt:81,AnRuZa:05}, and for these cases the model is found to undergo a phase transition at some inverse temperature $\beta_{\text{t}}(J)$ which is a continuously varying, nontrivial function of the coupling constant $J$. Similarly, for arbitrary on-site potentials $U$, a nontrivial dependence on $J$ of the transition (inverse) temperature $\beta_{\text{t}}$ as well as of the corresponding transition potential energy $v_{\text{t}}$ is expected (and corroborated by numerics).

From the point of view of configuration space topology, the class of solid-on-solid models \eqref{eq:SOS} is particularly simple. A complete characterization of the topology of the $\MM_v$ has been outlined by \textcite{Kastner:04}, and details are given in the Appendix.\footnote{This result includes, as a special case, those of \textcite{GriMo:04} for the Peyrard-Bishop model, obtained by a different method.} For any on-site potential $U$ which is a monotonous function on the intervals $\left(0,x_{\text{min}}\right)$ and $\left(x_{\text{min}},\infty\right)$ that are left and right to the location $x_{\text{min}}$ of the unique minimum of $U$, one finds that two topology changes take place within the family $\left\{\MM_v\right\}_{v\in\RR}$ of configuration space subsets. One is located at the ground state energy $v_1=U\left(x_{\text{min}}\right)$, while a second one appears at $v_2=\lim\limits_{x\to\infty}U(x)$. Both, $v_1$ and $v_2$ are {\em independent}\/ of the value of the coupling constant $J$. Comparing this finding with the nontrivial dependence of the transition potential energy $v_{\text{t}}$ as argued above, we observe that phase transitions in the class of models defined by Eq.\ \eqref{eq:SOS} are not related to topology changes in configuration space.

Setting $n=2$ in Eq.\ \eqref{eq:SOS}, the potential $V_\text{\tiny SOS}$ is smooth and the model fulfills, apart from the confining property, all requirements of Theorem~\ref{thrm:1}. We therefore conclude that the assumption in this theorem of the potential being confining cannot be relaxed in general.

\subsection{Nonanalyticities from maximization}
\label{sec:sing_max}
The examples presented in Secs.~\ref{sec:long-range} and \ref{sec:nonconfining} dem\-on\-strate that, at least for systems with long-range interactions and for systems with nonconfining potentials, a phase transition is not necessarily accompanied by a topology change in configuration space. In these cases one might suspect, instead of a topology change, a {\em different}\/ kind of mechanism behind the occurrence of a phase transition. In this section we argue that for both, the cases of long-range systems and nonconfining potentials discussed in Secs.~\ref{sec:long-range} and \ref{sec:nonconfining}, nonanalyticities in thermodynamic functions are generated from smooth functions by means of a maximization mechanism \cite{Kastner2:06}.

\subsubsection{Mean-field $\varphi^4$ model revisited}
\label{sec:sing_max_phi4}
The long rangedness of the pair potential of a system has, as mentioned in Sec.~\ref{sec:long-range}, remarkable consequences on the convexity properties of the thermodynamic functions. Notably, the (configurational) microcanonical entropy of a system with long-range interactions, in contrast to the short-range case \cite{Lanford:73,Gallavotti}, is not necessarily concave \cite{Dauxois_etal,TouElTur:04}.
\begin{defn}
\renewcommand{\labelenumi}{(\roman{enumi})}
\begin{enumerate}
\item A set $A\subseteq\RR^n$ is called a {\em convex set}\/ if $ax+(1-a)y\in A$ for all $x,y\in A$, $a\in[0,1]$.
\item A function $f:A\to\RR$ defined on a convex set $A$ is called a {\em convex function}\/ if
\begin{equation}
f(ax+(1-a)y)\leqslant af(x)+(1-a)f(y).
\end{equation}
\item If $-f$ is convex, $f$ is called a {\em concave function}.
\end{enumerate}
\end{defn}
In the case of long-range interactions, a nonanalytic point of the configurational microcanonical entropy $s_\infty(v)$ can arise from the maximization over one variable of a {\em smooth, but nonconcave}\/ entropy function $\tilde{s}_\infty(v,m)$ of {\em two}\/ variables (defined shortly), and this is precisely what happens for the mean-field $\varphi^4$ model.

For the mean-field $\varphi^4$ model defined by the potential \eqref{eq:potphi4}, an exact calculation of thermodynamic functions is possible, and large deviation techniques are an elegant way to perform such a calculation. \textcite{HaKa:05,HaKa:06} reported exact results for two related microcanonical thermodynamic functions of the mean-field $\varphi^4$ model in the thermodynamic limit: for the configurational microcanonical entropy $s_\infty(v)$ as defined in Eqs.\ \eqref{eq:s_conf} and \eqref{eq:sinf}, and for the configurational microcanonical entropy
\begin{multline}\label{eq:svm}
\tilde{s}_\infty(v,m)\\
=\lim_{N\to\infty}\frac{1}{N}\ln \int_{\Gamma_N} \dd q \Dirac\left[V(q)-Nv\right] \Dirac\big[\sum_{i=1}^N q_i-Nm\big]
\end{multline}
as a function of two variables, namely the potential energy $v$ and the magnetization $m$. The entropy $\tilde{s}_\infty$ is found to have a nonconcave part \cite{HaKa:05}. Furthermore, as a consequence of a general result from large deviation theory [Theorem I.4 in \textcite{Hollander}], $\tilde{s}_\infty$ is a smooth function in both its variables. Applying Laplace's method for the evaluation of asymptotic integrals to Eq.\ \eqref{eq:svm}, one can show that the entropy functions $s_\infty(v)$ and $\tilde{s}_\infty(v,m)$ are related by
\begin{equation}\label{eq:svm_sv}
s_\infty(v)
= \max_m\left[ \tilde{s}_\infty(v,m) \right].
\end{equation}
The maximization over $m$ of a smooth and nonconcave entropy function $\tilde{s}_\infty(v,m)$ then may lead to a nonanalyticity in $s_\infty$.

\begin{figure}[htb]
\center
\psfrag{0.1}{\small $0.1$}
\psfrag{0.2}{\small $0.2$}
\psfrag{0.3}{\small $0.3$}
\psfrag{0.4}{\small $0.4$}
\psfrag{0.5}{\small $0.5$}
\psfrag{-0.5}{\small $-0.5$}
\psfrag{-1.0}{\small $-1.0$}
\psfrag{0.0}{\small $0.0$}
\psfrag{1.0}{\small $1.0$}
\psfrag{0}{\small $0$}
\psfrag{-1}{\small $-1$}
\psfrag{-2}{\small $-2$}
\psfrag{-4}{\small $-4$}
\psfrag{v}{\small $v$}
\psfrag{m}{\small $m$}
\includegraphics[width=6cm]{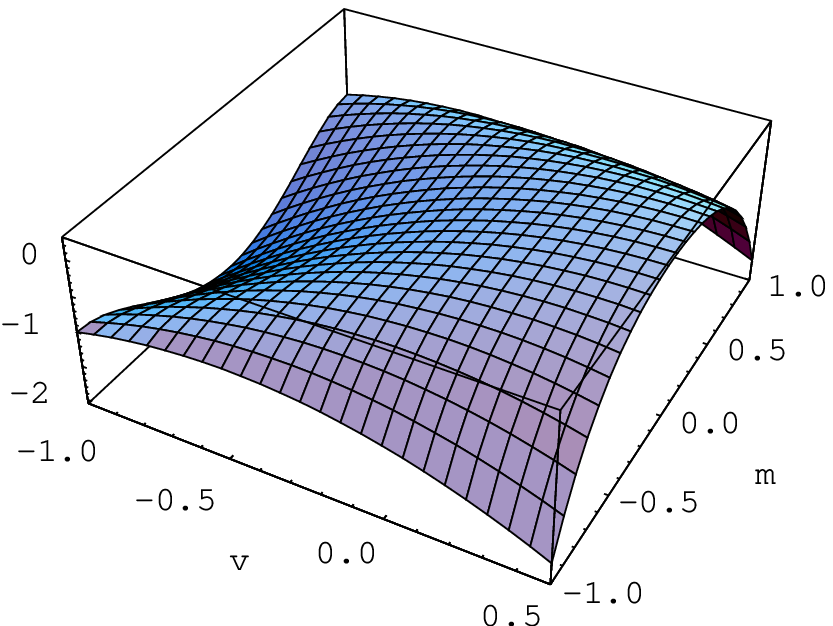}
\vspace{4mm}\hbox{}
\includegraphics[width=6cm]{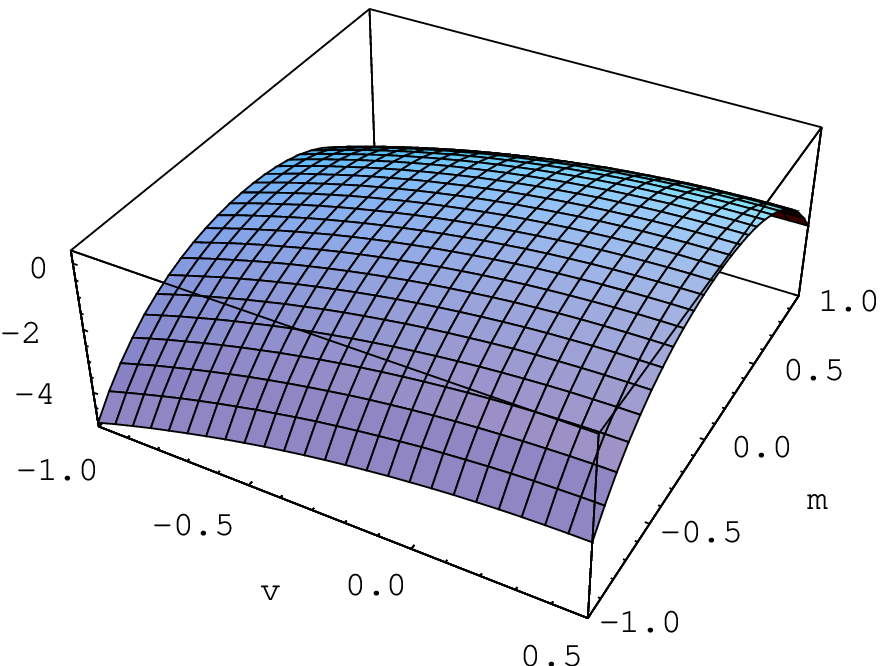}
\caption{\label{fig:nonconcave_s}
(Color online) Plots of the graphs of the nonconcave function $\bar{g}_1(v,m)$ (top) and of the concave function $\bar{g}_2(v,m)$ (bottom).}
\end{figure}

To illustrate how a nonanalyticity may emerge from a maximization over one variable of a smooth but nonconcave function, it is instructive to consider the following simple example. The function $\bar{g}_1:\RR^2\to\RR$ with
\begin{equation}
\bar{g}_1(v,m)=v-v^2-2vm^2-m^4
\end{equation}
is, as is easily verified, a nonconcave function, and its graph is shown in the plot of Fig.~\ref{fig:nonconcave_s} (top). Maximizing with respect to the second variable of $\bar{g}_1$, one obtains the function
\begin{equation}
g_1(v)=\max_m \bar{g}_1(v,m) = \begin{cases}v & \text{for $v<0$},\\ v-v^2 & \text{for $v\geqslant 0$},\end{cases}
\end{equation}
which has a nonanalytic point at $v=0$. In contrast, for a concave function like
\begin{equation}
\bar{g}_2(v,m)=v-v^2-2m^2-m^4
\end{equation}
[Fig.~\ref{fig:nonconcave_s} (bottom)], a maximization with respect to $m$ yields
\begin{equation}
g_2(v)=\max_m \bar{g}_2(v,m) = v-v^2,
\end{equation}
which is smooth on $\RR$.

\begin{figure}[ht]
\center
\psfrag{0}{\small $0$}
\psfrag{1}{\small $1$}
\psfrag{2}{\small $2$}
\psfrag{-1}{\small $-1$}
\psfrag{-2}{\small $-2$}
\psfrag{0.0}{\small $0.0$}
\psfrag{-0.5}{\small $-0.5$}
\psfrag{-1.0}{\small $-1.0$}
\psfrag{v}{\small $v$}
\psfrag{m}{\small $m$}
\includegraphics[width=7cm]{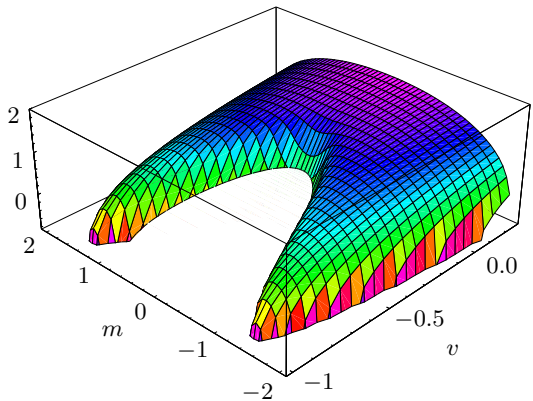}
\caption{\label{fig:smv3dplot}
(Color online) Plot of the graph of the entropy function $s_\infty(v,m)$ of the mean-field $\varphi^4$ model with coupling constant $J=1$. Adapted from \textcite{HaKa:06}.}
\end{figure}

Similar to the simple examples of the functions $\bar{g}_1$ and $g_1$, the nonanalytic point of the entropy $s_\infty$ of the mean-field $\varphi^4$\nobreakdash-model is created from a smooth but nonconcave entropy $s_\infty(v,m)$ by a maximization with respect to $m$ (for a plot of the graph of $s_\infty(v,m)$, see Fig.~\ref{fig:smv3dplot}). We interprete this maximization as the nonanalyticity-generating mechanism which is at the basis of the phase transition of the model.

In the thermodynamic limit (if it exists), the mean-field approach is exact for many systems with long-range interactions [see \textcite{BouBa:05} and references therein for a detailed discussion]. Since the mean-field variational problem can be written as a maximization of a smooth many-variable function, the same can be done for the entropy of these long-range systems. The different types of phase transitions which may result from such a maximization depend on the number of variables of the entropy function and on the symmetries present. A complete classification of the various nonanalyticities which can occur for an entropy function of two variables has been worked out by \textcite{BouBa:05}. 

In  models with short-range interactions, the above-described mechanism cannot occur due to the concavity \cite{Lanford:73,Gallavotti} of entropy functions. This is the content of the following theorem.
\begin{thrm}
Let $s^{(n)}(x_1,\dotsc,x_n)$ be a smooth, concave entropy function of $n$ variables. Then the corresponding reduced entropy function of $n-1$ variables,
\begin{multline}
s^{(n-1)}(x_1,\dotsc,x_{i-1},x_{i+1},\dotsc,x_n)\\
=\max_{x_i}\bigl[s^{(n)}(x_1,\dotsc,x_n)\bigr],
\end{multline}
is again smooth and concave.
\end{thrm}
\begin{proof}
By elementary calculus or geometric considerations.
\end{proof}
For systems with short-range interactions, this result rules out the occurrence of nonanalyticities from the above described maximization mechanism: If $s^{(n)}$ is smooth, then, as a consequence of its concavity, $s^{(n-1)}$ has to be smooth as well and no phase transition takes place. Nonetheless, in short-range models a different type of max\-i\-mi\-za\-tion can be the origin of a nonanalyticity, and this is the content of the following section.

\subsubsection{Solid-on-solid models revisited}
\label{sec:sing_max_SOS}
For the solid-on-solid models defined by potential \eqref{eq:SOS}, the configurational canonical free energy $f_\infty$ is expressed in Eq.~\eqref{eq:EV_max} as the logarithm of the largest of the eigenvalues $\lambda_i$ of the eigenvalue problem \eqref{eq:transfer},
\begin{equation}\label{eq:EV_max2}
-\beta f_\infty\left(\beta\right)=\max_{i}\left[\ln\lambda_i(\beta)\right],
\end{equation}
and the nonanalyticities of $f_\infty$ are discussed by studying the behavior of $\lambda_i$. The $\lambda_i$ are expected to be smooth functions of $\beta$, but they may have crossing points.\footnote{Related results exist in the perturbation theory of linear operators. Modifying those to turn this claim into a rigorous result would be a worthwhile task.} If we assume this claim to be correct, the maximization over the index $i$ in Eq.\ \eqref{eq:EV_max2} in the presence of a crossing point of the largest and the second largest $\lambda_i$ is the only remaining source of a nonanalyticity in $f_\infty$. For short-range models like the solid-on-solid models discussed here, equivalence of statistical ensembles holds. As a consequence, the entropy $s_\infty$ can be obtained from the free energy $f_\infty$ by means of a Legendre-Fenchel transform, and the nonanalytic point in $f_\infty$ gives rise to a nonanalytic point in $s_\infty$.

\subsection{Two nonanalyticity generating mechanisms}
Two counterexamples disproving the general validity of Conjectures~\ref{con:strong} and \ref{con:weak} on the relation between phase transitions and configuration space topology were presented in Secs.~\ref{sec:long-range} and \ref{sec:nonconfining}. For these models---the mean-field $\varphi^4$ model and a solid-on-solid model---we have shown that nonanalyticities in the thermodynamic functions may be viewed as arising from the maximization over one variable (or one discrete index) of some smooth function. This maximization can be interpreted as one possible mechanism generating a nonanalyticity, whereas (certain) topology changes in the configuration space subsets $\MM_v$ are another such mechanism. In principle, in a given system any of these mechanism, or even both, may occur and trigger the occurrence of a phase transition. Theorem~\ref{thrm:1} then asserts that, for the class of (short-range, non-confining, etc.) systems fulfilling its assumptions, topology changes are the {\em only}\/ nonanalyticity generating mechanism at ones disposal.

The type of nonanalyticity generating mechanism occurring has notable consequences also for physically relevant quantities: In case of a continuous phase transition, one can show that generically\footnote{Meaning ``whenever none of the leading coefficients of a Taylor expansion of $\tilde{s}$ in the vicinity of the phase transition point accidentally vanishes.''} the critical exponents characterizing the nonanalytic point take on mean-field values whenever the nonanalyticity was created by a maximization as in Eq.\ \eqref{eq:svm_sv} [see Appendix of \textcite{HaKa:05} for details].

The identification of a second nonanalyticity generating mechanism does by no means diminish the interest in the topological approach to phase transitions. The typical systems of interest in statistical mechanics have short-range interactions and confining potentials, and an investigation of the relation of phase transitions and configuration space topology in such systems is of great interest. However, for this class of systems analytic calculations of topological quantities seem out of reach, which underlines the importance in further development of numerical techniques as in Sec.~\ref{sec:numerics}. 

\section{Search for a sufficiency criterion}
\label{sec:sufficiency}
We turn our attention back to the class of short-range systems for which, according to Theorem~\ref{thrm:1}, a topology change within the family $\left\{\MM_v\right\}_{v\in\RR}$ of configuration space subsets is {\em necessary}\/ for a phase transition to occur at the corresponding energy or temperature. Although this theorem indicates that some sort of relation between phase transitions and configuration space topology exists, it does not have much to say about the {\em form}\/ of this relation. As pointed out in Sec.~\ref{sec:theorems}, a topology change of $\MM_v$ is not sufficient for a phase transition to take place, and the obvious question to ask is:

{\em Under which conditions do topology changes give rise to a phase transition?}

This search for a sufficiency criterion, specifying the relation between phase transitions and topology changes, may be considered as {\em the}\/ big open question in the field, and from an answer to this question one can expect to gain insights into the fundamental mechanisms which are at the origin of a phase transition. No final answer to this question has been given so far, but some of the preliminary results are worth mentioning. The following sections discuss several proposals which have been suggested, more or less explicitly, on the basis of the few model calculations which are available.

\subsection{Simultaneous attachment of $\Order(N)$ different handles}
\label{sec:ON_handles}
\textcite{CaPeCo:03} computed critical points and indices of the configuration space subsets $\MM_v$ of the mean-field $XY$ model with and without an external magnetic field. This model is characterized by the potential
\begin{equation}
V_{XY}(q)=\frac{J}{2N}\sum_{i,j=1}^N\left[1-\cos(q_i-q_j)\right]-h\sum_{i=1}^N\cos q_i,
\end{equation}
where $J>0$ is a coupling constant, $h\in\RR$ is an external magnetic field, and the coordinates $q_i\in[0,2\pi)$ are angular variables. The potential energy $v_{\text{t}}$ at which a phase transition occurs in this model for $h=0$ is found to coincide with the only critical value at which the topology of the $\MM_v$ involves the simultaneous attachment of handles\footnote{See Theorem~\ref{thm:handle_decomposition} of the present paper or Sec.~3 of \textcite{Matsumoto} for the definition of a handle in topology and for the construction of manifolds by the attachment of handles.} of $\Order(N)$ different types. This observation led \textcite{CaPeCo:03} to the following conjecture.
\begin{conjecture}\label{con:ON_handles}
A topology change of $\MM_v$ at some $v=v_{\text{t}}$ which involves the simultaneous attachment of handles of $\Order(N)$ different types entails a phase transition at $v_{\text{t}}$.
\end{conjecture}
However, a counterexample to this conjecture can be constructed by considering the mean-field $XY$ model in the presence of an external magnetic field $h$. In this case, the model does not show a phase transition, but the same type of topology changes as in the absence of a field $h$.

\subsection{Nonanalyticities of the Euler characteristic}
In the same paper \cite{CaPeCo:03}, a nonanalytic point in the logarithmic density of the Euler characteristic $\chi(\MM_v)$ of the mean-field $XY$ model at $v=v_\text{t}$ is reported, and the same feature is observed for the mean-field $k$-trigonometric model (see Sec.~\ref{sec:ktrig} and Fig.~\ref{fig:ktrig}). These findings suggest the following conjecture.
\begin{conjecture}\label{con:Eulerkink}
A nonanalyticity at $v=v_{\text{t}}$ in the (logarithmic density of the) Euler characteristic of $\MM_v$ entails a phase transition at $v_{\text{t}}$.
\end{conjecture}
This is a ``sufficiency version'' of Conjecture~\ref{con:strong}, but it suffers from the same shortcoming as the---presumably related---Conjecture~\ref{con:ON_handles}: For nonzero external magnetic field, the nonanalyticity in the Euler characteristic of the mean-field $XY$ model persists, although no phase transition is present. 

\subsection{Nonpurely topological sufficiency conditions}
For two different models (i.\,e., with and without external magnetic field), we have seen that in the presence of the same kind of topology changes the thermodynamic properties can differ drastically, and this behavior can be understood as follows [see \textcite{CaKa:07} for a related discussion].

\begin{figure}[htb]
\center
\psfrag{0.1}{$0.1$}
\psfrag{0.2}{$0.2$}
\psfrag{0.3}{$0.3$}
\psfrag{0.4}{$0.4$}
\psfrag{0.5}{$0.5$}
\psfrag{-0.5}{$-0.5$}
\psfrag{-1.0}{$-1.0$}
\psfrag{-1.5}{$-1.5$}
\psfrag{-2.0}{$-2.0$}
\psfrag{-2.5}{$-2.5$}
\psfrag{-0.4}{$-0.4$}
\psfrag{-0.2}{$-0.2$}
\psfrag{0.6}{$0.6$}
\psfrag{v}{$v$}
\psfrag{s}{$s_\infty$}
\includegraphics[height=4.5cm]{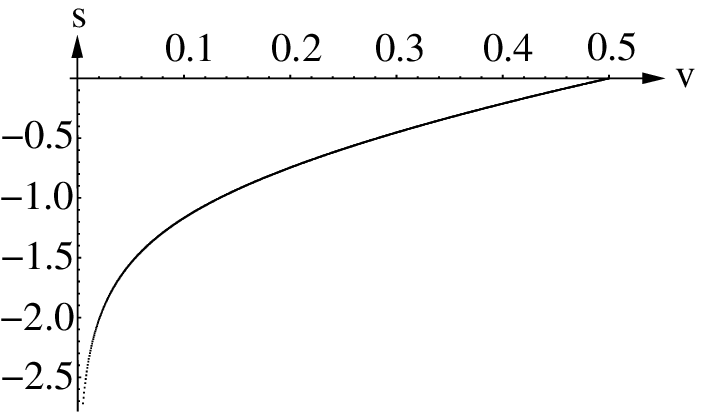}
\vspace{1mm}\hbox{}
\includegraphics[height=4.5cm]{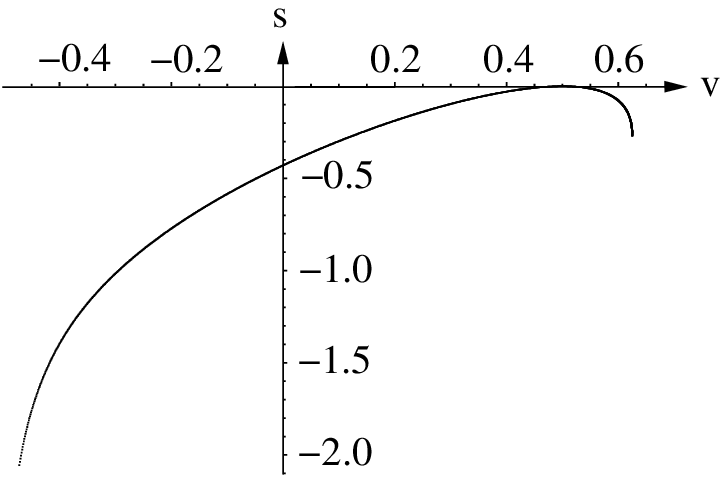}
\caption{\label{fig:sXY}
Configurational microcanonical entropy $s_\infty(v)$ of the mean-field $XY$ model in the thermodynamic limit. In the upper plot, which is for zero external magnetic field, the slope of $s_\infty$ has a positive lower bound. For an external magnetic field $h=1/2$, the graph of $s_\infty$ is shown in the lower plot, and the slope of $s_\infty$ is unbounded above and below. Adapted from \textcite{CaKa:07}.}
\end{figure}

The potential $V_{XY}/N$ of the mean-field $XY$ model is a bounded above and below function, and the value $v_\text{t}$ at which $\Order(N)$ different handles are attached simultaneously (or at which the Euler characteristic has a nonanalytic point) equals its upper bound,
\begin{equation}
v_\text{t}=\frac{1}{N}\sup_{q\in\Gamma_N}V_{XY}(q).
\end{equation}
In the absence of an external magnetic field, the configurational microcanonical entropy $s_N$ is a monotonously increasing function, and its slope has a positive lower bound,
\begin{equation}
\inf_v\frac{\partial s_N(v)}{\partial v}>0
\end{equation}
[see Fig.~\ref{fig:sXY} (top) for a plot of the graph of $s_\infty$. The point $v_\text{t}$ at which $\Order(N)$ different handles are attached corresponds to the ``end point'' of $s_N$ at the upper boundary of its domain. Switching, via Eq.~\eqref{eq:Legendre}, to the canonical ensemble by considering the Legendre-Fenchel transform of $s_N$, this end point is conjugate to the value $\beta_\text{t}=\inf_v[\partial s_N(v)/\partial v]$ of the inverse temperature, and it is at this value that a phase transition occurs in the mean-field $XY$ model.

In contrast, for nonzero external field $h$, even if arbitrarily small, the shape of $s_N$ changes drastically. The graph of $s_N$ then looks as plotted in Fig.~\ref{fig:sXY} (bottom), and its slope $\partial s_N(v)/\partial v$ is not anymore bounded below. The value $v_\text{t}$ at which $\Order(N)$ different handles are attached corresponds to the maximum of $s_N$. Canonically, however, the temperature at $v_\text{t}$ is infinite
\begin{equation}
T_\text{t}=\frac{1}{\beta_\text{t}}=\left(\frac{\partial s_N(v)}{\partial v}\right)^{-1}\Big|_{v=v_\text{t}}=\infty.
\end{equation}
Therefore the corresponding macrostate is thermodynamically not accessible and the associated ``strong'' topology change does not affect the thermodynamic behavior of the system.

Taking into account the above considerations, a sufficiency criterion of {\em purely}\/ topological nature for the existence of a phase transition is unlikely to exist. ``Strong'' topology changes in the sense of conjecture \ref{con:ON_handles} or \ref{con:Eulerkink} [attachment of handles of $\Order(N)$ different types or a nonanalyticity of the Euler characteristic] are reasonable candidates for being part of a sufficiency criterion, but---as follows from the discussion of the mean-field $XY$ model with external magnetic field---apparently have to be supplemented by a nontopological condition (presumably comprising some notion of {\em measure} on phase space or configuration space).%
\footnote{A sufficiency criterion for the existence of a phase transition, consisting of a topological part {\em and}\/ a probabilistic part, has been given by \textcite{BaCa06} for discrete symmetry breaking phase transitions. The probabilistic ingredient of their theorem is a Peierls-type argument, and the phase transition to which the criterion refers does in general not take place at the energy of the topology change considered, so---although making reference to topology in configuration space---the criterion in \textcite{BaCa06} is not quite in the spirit of the topological criteria discussed before.}
Note, however, that in general the influence of ``measure'' may be of a more subtle kind than the ``infinite-temperature''-argument applying to the mean-field $XY$ model.

\section{Summary}
\label{sec:conclusions}
We have reviewed and critically discussed a topological approach to phase transitions, investigating the relation of phase transitions in classical statistical mechanical systems and topology changes of the configuration space subsets $\MM_v$ as defined in Eq.\ \eqref{eq:Mv}. For the computation of the topology of the $\MM_v$, the mathematical framework of Morse theory is particularly convenient, and we have summarized some of its elementary results in view of this application.

In finite systems, topology changes of $\MM_v$ are intimately related to nonanalytic points of the configurational microcanonical entropy $s_N$. Any critical point of a potential $V$ gives rise to a nonanalytic point of $s_N(v)$ at $v=V(q_\text{c})/N$. The form of the nonanalyticity is specified in Theorem \ref{thm:finite}, and the order $n$ of the derivative for which $\partial^n s_N/\partial v^n$ becomes discontinuous increases linearly with $N$.

In the thermodynamic limit, the relation between nonanalytic points of thermodynamic functions (i.\,e., phase transitions) and topology changes of $\MM_v$ is more intricate. For some class of short-range systems with smooth, nonconfining, and bounded below potentials, a topology change of the subsets $\MM_v$ at $v=v_{\text{t}}$ is necessary, but not sufficient for a phase transition to take place at $v_{\text{t}}$. In contrast, in systems with long-range interactions or systems with nonconfining potentials a phase transition need not be accompanied by such a topology change, as demonstrated by means of two counterexamples: the mean-field (and therefore long-ranged) $\varphi^4$ model and some class of solid-on-solid models with nonconfining potentials. For such systems, the nonanalytic point in a thermodynamic function can be viewed as emerging from a maximization over some smooth function.

In summary, two different mechanisms which may cause a nonanalyticity in a thermodynamic function have been identified: first, certain topology changes within the family $\left\{\MM_v\right\}_{v\in\RR}$ of configuration space subsets; and, second, a maximization over one variable of a smooth function of several variables. Theorem~\ref{thrm:1} then asserts that only the former one of these mechanism can occur in the class of short-range systems which are in accordance with the theorem's assumptions. This is a remarkable finding, since this class of systems contains the types of the systems which are typically of interest in statistical physics.

It remains an open task to precisely specify which topology changes entail a phase transition. Several proposals for conditions on topology changes of the $\MM_v$, allegedly sufficient to guarantee the occurrence of a phase transition, are discussed, but a final answer to this question is still lacking. One may conjecture that such a criterion will not be exclusively of topological character, but instead may involve some notion of measure or geometry as well. A solution to this problem will be a major step forward towards an understanding of the origin of phase transitions in classical statistical mechanical systems.

\section{Epilog}
From the above discussion, it should be obvious to the reader that the relation of phase transitions to topology changes in configuration space is not a settled issue, but a topic of current research activity. Despite its incomplete status, one may profit from the study of the results to date in various ways.

First, the topological approach stimulates the study of phase transitions from a viewpoint quite different from the conventional one, and such a change of perspective may help to deepen the understanding and to inspire further research activity. In particular for the study of the relation between phase transitions and the chaoticity of the underlying dynamics of the system, the topological viewpoint may be beneficial \cite{CaPeCo:00}. Again from a conceptual point of view, the topological approach, at least within the framework of Morse theory as used throughout the present paper, has remarkable similarities to the study of glassy systems, biomolecules, or clusters from the saddle points of their potential energy landscapes \cite{Wales}, and one may hope to profit from these similarities of methods in future investigations.

Apart from such general and conceptual considerations, early efforts have been seen to bring to fruit differential geometrical concepts, and possibly also the related topological concepts, in physics applications. \textcite{MazCa:06} considered curvature fluctuations of constant potential energy submanifolds of minimalistic models of proteins, finding that good folders may be distinguished from bad folders by studying the energy dependence of these fluctuations. A connection between curvature fluctuations and topology changes in configuration space as described by \textcite{CaPeCo:00} might then lead to a topological interpretation of this result.

Last but not least, presenting and advocating a concept applicable to classical statistical mechanical systems inevitably provokes the question of how an extension to quantum mechanical systems might look. The obvious idea would be to study the topology of constant energy subsets of the underlying Hilbert space (or of some related projective space), but to the knowledge of the author no noteworthy effort along these lines has been made so far. What may be of related interest is a study of the topology of constant energy surfaces in the complex projective space of pure quantum states reported by \textcite{BroHugh:01} and \textcite{BroHoHu:07} for quantum mechanical one-particle systems, and one might take these results as a starting point for a future investigation of phase transitions in quantum systems and their relation to state space topology.

\begin{acknowledgments}
I would like to thank Helmut B\"uttner, Lapo Casetti, and Hugo Touchette for helpful comments.
\end{acknowledgments}

\appendix
\section{Configuration space topology of solid-on-solid models}
\label{app:SOS-topology}
In this Appendix, a theorem characterizing the topology of the configuration space subsets $\MM_v$ for the class of solid-on-solid models \eqref{eq:SOS} defined and discussed in Sec.~\ref{sec:nonconfining} is proven.
\begin{thrm}\label{thrm:SOS}
Consider the potential
\begin{equation}\label{eq:SOS_app}
V(q)=J\sum_{i=1}^{N-1} \left|q_{i+1}-q_i\right|^n + \sum_{i=1}^N U\left(q_i\right)
\end{equation}
with $J>0$, $n\in\NNN$, and $q_i\in\RR^+$ for all $i=1,\dotsc,N$. The on-site potential $U:\RR^+\to\RR$ is supposed to have a unique minimum for some argument $x_{\text{min}}$, and we define $U_{\text{min}}=U\left(x_{\text{min}}\right)$. Let $U$ be monotonously decreasing on the interval $\left(0,x_{\text{min}}\right)$ and monotonously increasing on $\left(x_{\text{min}},\infty\right)$. Furthermore, let $\lim\limits_{x\to\infty}U(x)=U_\infty<\infty$, which makes $V$ a nonconfining potential. Under these conditions, the subsets $\MM_v=\big\{q\in\left(\RR^+\right)^N\,\big|\,V\left(q\big)\leqslant vN\right\}$ fulfill the equivalence relations
\begin{equation}\label{eq:MvSOS}
\MM_v \sim
\begin{cases}
\emptyset & \text{for $v< U_{\text{min}}$},\\
{\mathbbm I}^N & \text{for $U_{\text{min}} < v < U_\infty$},\\
\RR^+\times{\mathbbm I}^{N-1} & \text{for $U_\infty < v$},
\end{cases}
\end{equation}
where $\sim$ denotes homeomorphicity, $\emptyset$ is the empty set, and ${\mathbbm I}=[0,1]$ the unit interval.
\end{thrm}
\begin{proof}
In a first step, $\MM_v$, if not empty, is shown to be a {\em star convex}\/ subset of $\left(\RR^+\right)^N$, i.\,e., there exists a $\widetilde{q} \in \MM_v$ such that the line segment from $\widetilde{q}$ to any point in $\MM_v$ is contained in $\MM_v$. This is proven by observing that, for every $q=(q_1,...,q_N)\in\left(\RR^+\right)^N$ and every $\lambda\in[0,1]$, the inequality
\begin{equation}\label{eq:starconvex}
\begin{split}
&\shoveleft V\left(\lambda \left(q-q_{\text{min}}\right)\right)\\
&= \sum_{i=1}^{N-1} \underbrace{\lambda^n J\left|q_{i+1}-q_i\right|^n}_{\displaystyle \leqslant J\left|q_{i+1}-q_i\right|^n} + \sum_{i=1}^N \underbrace{U\left(\lambda \left(q_i-x_{\text{min}}\right)\right)}_{\displaystyle \leqslant U\left(q_i-x_{\text{min}}\right)}\\
&\leqslant V\left(q-q_{\text{min}}\right)
\end{split}
\end{equation}
holds. Here $q_{\text{min}}=\left(x_{\text{min}},\dotsc,x_{\text{min}}\right)$, and the inequality for the second term in Eq.\ \eqref{eq:starconvex} is a consequence of the monotonicity properties of $U$. The star convexity of $\MM_v$ implies {\em homotopical}\/ equivalence to ${\mathbbm I}^N$ (or to an $N$-ball ${\mathbbm B}^N$), but not necessarily homeomorphicity.

In a second step, the {\em (un)boundedness}\/ of $\MM_v$ is investigated. This is done, analogously to the treatment by \textcite{GriMo:04}, by studying the asymptotic behavior of $V(\lambda q)$ in the limit $\lambda\to\infty$. As a consequence of the nonconfining character of $U$, we find
\begin{equation}
\lim_{\lambda\to\infty} V(\lambda q) =\begin{cases}NU_\infty<\infty & \text{if $q_i=q_j\;\forall i,j=1,\dotsc,N$},\\
\infty & \text{else}.
\end{cases}
\end{equation}
Hence for $v=V/N<U_\infty$, only configurations $q\in \left(\RR^+\right)^N$ with finite (Euclidean) norm are accessible, whereas configurations of arbitrarily large norm can be attained for $v>U_\infty$. From this observation it follows that $\MM_v$ is a bounded subset of $\left(\RR^+\right)^N$ for $v<U_\infty$, and, together with the star convexity shown above, $\MM_v$ is found to be homeomorphic to ${\mathbbm I}^{N}$. For $v>U_\infty$, however, $\MM_v$ is unbounded. Since configurations of arbitrarily large norm can be attained only ``in a {\em single}\/ spatial direction'', i.\,e., in the vicinity of the (hyper)space diagonal $q=\lambda(1,...,1)$, $\lambda>0$, we conclude that $M_v^b$ is topologically equivalent to the product $M_v^b\sim\RR^+\times {\mathbbm I}^{N-1}$. With the immediate observation that $\MM_v=\emptyset$ for $v<U_{\text{min}}$, the proof of Theorem~\ref{thrm:SOS} is complete.
\end{proof}

\bibliographystyle{apsrmplong}
\bibliography{TopoPT}

\end{document}